\documentclass[12pt]{article}

\usepackage{epsfig}
\usepackage{latexsym}
\def\beq{\begin{equation}}
\def\eeq#1{\label{#1}\end{equation}}
\def\eeqn{\end{equation}}

\def\beqa{\begin{eqnarray}}
\def\eeqa#1{\label{#1}\end{eqnarray}}
\def\eeqan{\end{eqnarray}}

\let\bar=\overbar

\def\Dslash{\not{\hbox{\kern-4pt $D$}}}
\def\dslash{\not{\hbox{\kern-2pt $\del$}}}

\def\msb{{\bar{\ssstyle M \kern -1pt S}}}

\newcommand {\pom} {I\!\!P}
\newcommand {\pomsub} {{\scriptscriptstyle \pom}}

\newcommand {\xpom} {x_{\pomsub}}
\newcommand {\apom} {\alpha_{\pomsub}}

\newcommand {\aprime} {\alpha^\prime_\pomsub}

\newcommand{\gv}{\gamma^\star}
\newcommand{\gp}{\gamma p}
\newcommand{\gvp}{\gamma^\star p}
\newcommand\units{\,\mathrm}
\newcommand{\gev}{\units{GeV}}
\newcommand{\gevtwo}{\units{GeV^2}}
\newcommand{\gevmtwo}{\units{GeV^{-2}}}

\newcommand{\ftwod}{F_2^D}
\newcommand{\ftwodthree}{F_2^{D(3)}}
\newcommand{\ftwodfour}{F_2^{D(4)}}

\newcommand{\ftwopom}{F_2^{\pomsub}}

\def\Title#1{\begin{center} {\Large {\bf #1} } \end{center}}

\begin{document}

\Title{Diffraction and the Pomeron}

\bigskip\bigskip


\begin{raggedright}  

{\it H. Abramowicz\index{Abramowicz, H.}\\
School of Physics and Astronomy\\
Raymond and Beverly Sackler Faculty of Exact Sciences\\ 
Tel Aviv University, Israel }
\bigskip\bigskip
\end{raggedright}

\section{Introduction}

In hadron-hadron scattering, interactions are classified by the
characteristics of the final states.  In elastic scattering, both
hadrons emerge unscathed and no other particles are produced. In
diffractive scattering, the energy transfer between the two
interacting hadrons remains small, but one (single dissociation) or
both (double dissociation) hadrons dissociate into multi-particle
final states, preserving the quantum numbers of the associated initial
hadron.  The remaining configurations correspond to inelastic interactions.

The first interpretation of diffraction, due to Good and
Walker~\cite{Good:1960}, \index{Good, M. L.} \index{Walker, W. D.} was
that different components of the projectile were differently absorbed
by the target, leading to the creation of new physical states. This was
the first indication for the composite nature of hadrons.

In the Regge theory of strong interactions~\cite{Collins:1977},
\index{Regge theory} diffraction is the result of exchanging a universal
trajectory with the quantum numbers of the vacuum, the (soft) Pomeron, $\pom$,
introduced by Gribov~\cite{Gribov:1961}. \index{Gribov, V.  N.}
\index{Pomeron} 

In the language of Quantum Chromodynamics, the candidate for vacuum
exchange with properties similar to the soft Pomeron is two gluon
exchange~\cite{Low:1975,Nussinov:1975}. \index{Low, F E}
\index{Nussinov, S} As a result of interactions between the two
gluons, a ladder structure develops. In perturbative QCD, the
properties of this ladder depend on the energy and scales involved in
the interaction, implying its non-universal character. In the
high-energy limit, the properties of the ladder have been derived for
multi-Regge kinematics and the resulting exchange is called the (hard)
BFKL pomeron~\cite{Lipatov:1976,Kuraev:1977,Balitsky:1978}.
\index{BFKL pomeron}

A renewed interest in diffractive scattering followed the observation
of a copious production of diffractive-like events in deep inelastic
scattering (DIS) at the HERA $ep$ collider~\cite{ZEUSdiff1,H1diff1} as
well as the earlier observation of jet production associated with a
leading proton in $p\bar{p}$ at CERN~\cite{UA8}. The presence of a
large scale opens the possibility of studying the partonic structure
of the diffractive exchange as suggested by Ingelman and
Schlein~\cite{Ingelman:1985} \index{Ingelman} \index{Schlein} and
testing QCD dynamics.  Moreover, the study of diffractive scattering
offers a unique opportunity to understand the relation between the
fundamental degrees of freedom prevailing in soft interactions --
hadrons and Regge trajectories, and those of QCD -- quarks and gluons.
One of the challenges is to establish theoretically and experimentally
the reactions in which the soft component is dominant and those in
which the perturbative QCD formalism is applicable. In this report the
discussion will focus on single diffractive processes in the presence
of a large scale, mainly studied at HERA and at FNAL.

\section{Kinematics of diffractive scattering}

The variables used to analyze diffractive scattering will be
introduced for $ep$ DIS. Since DIS is perceived as a two-step process,
in which the incoming lepton emits a photon which then interacts with
the proton target, the relevant variables can be readily generalized
to $p\bar{p}$ interactions.

\begin{figure}[htb]
\begin{center}
\epsfig{file=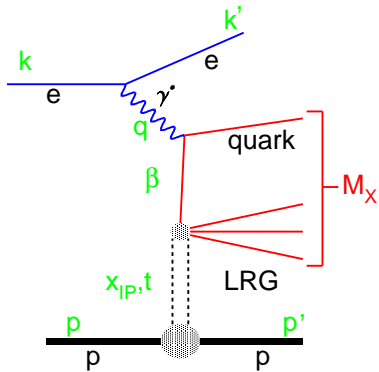,height=5cm}
\caption{Schematic diagram for diffractive DIS in $ep$ interactions.} 
\label{fig:dis-diag}
\end{center}
\end{figure}
A diagram for diffractive scattering \index{diffractive scattering} in
DIS, where the diffracted state is separated from the scattered proton
by a large rapidity gap (LRG), is presented in
Figure~\ref{fig:dis-diag} and all the relevant four vectors are
defined therein. In addition to the usual DIS variables,
$Q^2=-q^2=-(k-k^\prime)^2$, $W^2=(q+p)^2$, $x=\frac{Q^2}{2p\cdot q}$
and $y = \frac{p\cdot q}{p \cdot k}$, the variables used to
described the diffractive final state are,
\begin{eqnarray}
t &=& (p-p^\prime)^2 \, ,
\label{eq:deft} \\
\xpom &=& \frac{q\cdot (p-p^\prime)}{q \cdot p}
\simeq \frac{M_X^2+Q^2}{W^2+Q^2} \, ,
\label{eq:defxpom} \\
\beta &=& \frac{Q^2}{2q \cdot (P-P^\prime)} = \frac{x}{\xpom}
\simeq \frac{Q^2}{Q^2+M_X^2} \, .
\label{eq:defbeta}
\end{eqnarray}
$\xpom$ \index{\xpom} is the fractional loss of the proton
longitudinal momentum. It is sometimes denoted by $\xi$. $\beta$
\index{\beta} is the equivalent of Bjorken $x$ but relative to the
exchanged object. $M_X$ is the invariant mass of the hadronic final
state recoiling against the leading proton, $M_X^2=(q+p-p^\prime)^2$.
The approximate relations hold for small values of the four-momentum
transfer squared $t$ and large $W$, typical of high energy
diffraction.

\begin{figure}[htb]
\begin{center}
\epsfig{file=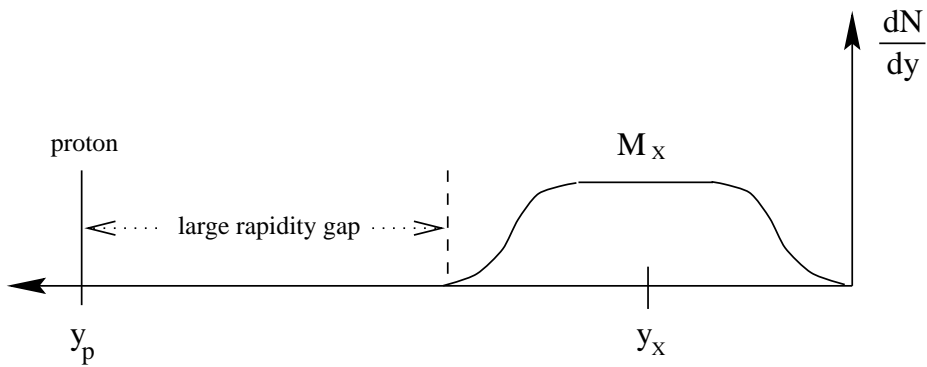,height=5cm} 
\caption{Schematic representation of rapidity ($y$) distribution  
for single diffraction events}  
\label{fig:lrg} 
\end{center}
\end{figure}

The need to preserve the identity of the target in diffractive
scattering limits the square of the momentum transfer, $|t| <
1/R_T^2$, where $R_T$ is the radius of the target. The $t$
distribution typically has an exponential behaviour, $f(t) \sim
\exp(-b|t|)$ with $b\simeq R_T^2/6$. The allowed $M_X$ is also limited
by the coherence requirement. The minimum value of $t$ required to
produce a given $M_X$ from a target with mass $m_T$ is $|t_{\mathrm
  min}| \simeq m_T^2 (M_X^2+Q^2)^2/W^4$. For a typical hadronic radius
of 1 fm, $M_X^2 <0.2 W^2$ and the hadronic final state exhibits a
large rapidity gap \index{large rapidity gap} between the fragments of
the diffracted state and the unscathed target (see
Figure~\ref{fig:lrg}).
Therefore in collider experiments, diffractive events are identified
either by the presence of a fast proton along the beam direction or by
the presence of a large rapidity gap in the central detectors.

\section{Formalism of diffractive scattering}

To describe diffractive DIS, it is customary to choose the variables
$\xpom$ and $t$ in addition to the usual $x$ and $Q^2$ in the cross
section formula. The four-fold differential cross section for $ep$
scattering can be written as
\begin{equation}
\frac{d^4\sigma^D_{ep}}{ d\,\xpom d\,t d\,x d\,Q^2 }
=\frac{2\pi \alpha^2}{x Q^4} \left[ 1+(1-y)^2\right] 
F_2^{D(4)}(x,Q^2,\xpom,t) \, ,\label{eq:f2d4} 
\end{equation}
where for the sake of simplicity the contribution from the
longitudinal structure function is omitted. The superscript $D$
denotes the diffractive process and the number in parenthesis in the
superscript of the $F_2^D(4)$ \index{diffractive structure function} is a
reminder that the units of the structure function have changed.
$\ftwodfour$ integrated over $t$ is denoted by $\ftwodthree$.

The structure function $F_2$ is related to the absorption cross
section of a virtual photon by the proton, $\sigma_{\gamma^\star p}$.
For diffractive scattering, in the limit of high $W$ (low $x$),
\begin{equation}
F_2^{D(4)}(x,Q^2,\xpom,t) = \frac{Q^2}{4\pi^2\alpha}
\frac{d^2\sigma^D_{\gamma^\star p}}{ d\,\xpom d\,t} \, .
\label{eq:gstarp}
\end{equation}
This relation allows predictions for diffractive scattering in DIS
based on Regge phenomenology applied to $\gvp$ scattering. In fact
many of the questions that are addressed in analyzing diffractive
scattering are inspired by Regge phenomenology as established in soft
hadron-hadron interactions.

\subsection{Regge phenomenology}

For scattering of two hadrons $a$ and $b$ at squared center of mass
energy $s \gg m^2_{a,b},\, t$, Regge phenomenology \index{Regge
  phenomenology} implies that,
\begin{eqnarray}
\sigma_{tot} (ab) &\sim& s^{\apom(0)-1} \, , 
 \label{eq:stot} \\
\frac{d^2\sigma^{el}}{dt}(ab \rightarrow ab) &=& 
\frac{\sigma^2_{tot}(ab)}{16\pi}e^{2 (b_0^{el} + \aprime \ln s) t} \, ,
\label{eq:sel} \\
\frac{d^2\sigma^D}{dt dM_X^2}(ab \rightarrow Xb) &\sim& \frac{1}{M_X^2} 
\left(\frac{s}{M_X^2}\right)^{2(\apom(0)-1)}
e^{2(b_0^{D}+\aprime \ln \frac{s}{M_X^2})t} \, , \label{eq:sdiff}
\end{eqnarray}
where $\apom(t)=\apom(0) + \aprime t$ is the parameterization of the
$\pom$ trajectory \index{Pomeron trajectory}.  The universality of
this parameterization has been pointed out by Donnachie and Landshoff.
The value of $\apom(0)=1.081$~\cite{Donnachie:1992} and $\aprime=0.25
\gevmtwo$~\cite{Donnachie:1984} were derived based on total
hadron-proton interaction cross sections and elastic proton-proton
data.  Recently the $\pom$ intercept has been
reevaluated~\cite{Kang:1998} leading to a value of $\apom(0)=1.096 \pm
0.03$. The $\pom$ intercept is sometimes presented as
$\apom(0)=1+\epsilon$.

Three implications are worth noting.\\
{\bf (1)} The slope of the $t$ distribution is increasing with $\ln
s$.  This fact, borne out by the data, is known as shrinkage
\index{shrinkage} of the $t$ distribution. It is due to the fact that
$\aprime \neq 0$ and has been explained by Gribov~\cite{Gribov:1961}
as diffusion of particles in the exchange towards low transverse
momenta, $k_T$, with $\aprime \sim 1/k_T^2$
(see also~\cite{Forshaw:1997}). \\
{\bf (2)} A steep and universal $\xpom$ dependence of the diffractive
cross section is expected, $d\sigma^D/d\xpom \sim \xpom^{-(1+2\epsilon)}$.\\
{\bf (3)} The ratios $\sigma^{el}/\sigma_{tot}$ and
$\sigma^{D}/\sigma_{tot}$ rise like $s^\epsilon$. Since $\epsilon>0$
this is bound to lead to unitarity violation.

\subsection{Perturbative QCD}

QCD factorization \index{QCD factorization} for the diffractive
structure function of the proton, $\ftwod$, is expected to
hold~\cite{Collins:1998,Berera:1996,Trentadue:1994}. $\ftwod$ is
decomposed into diffractive parton distributions \index{diffractive
  parton distributions}, $f^D_i$, in a way similar to the inclusive
$F_2$,
\begin{equation}
\frac{d\ftwod (x,Q^2,\xpom,t)}{d\xpom dt}=\sum_i \int_0^{\xpom} dz 
\frac{d f^D_i(z,\mu,\xpom,t)}{d\xpom dt} \hat{F}_{2,i}(\frac{x}{z},Q^2,\mu) \, ,
\label{eq:fact}
\end{equation}
where $\hat{F}_{2,i}$ is the universal structure function for DIS on
parton $i$, $\mu$ is the factorization scale at which $f^D_i$ are
probed and $z$ is the fraction of momentum of the proton carried by
the diffractive parton $i$. Diffractive partons are to be understood
as those which lead to a diffractive final state. The DGLAP evolution
equation applies in the same way as for the inclusive case. For a fixed
value of $\xpom$, the evolution in $x$ and $Q^2$ is equivalent to the
evolution in $\beta$ and $Q^2$.

Note that QCD factorization for the diffractive parton distributions
is expected to fail for hard diffraction in hadron
collisions~\cite{Collins:1998}.

If, following Ingelman and Schlein~\cite{Ingelman:1985}, one further
assumes the validity of Regge factorization, $\ftwod$ may be decomposed
into a universal $\pom$ flux and the structure function of $\pom$,
\begin{equation}
\frac{d\ftwod (x,Q^2,\xpom,t)}{d\xpom dt}= f_{\pom/p}(\xpom,t)
\ftwopom (\beta,Q^2) \, ,
\label{eq:f2pom}
\end{equation}
where the normalization of either of the two components is arbitrary.
It implies that the $\xpom$ and $t$ dependence of the diffractive
cross section is universal, independent of $Q^2$ and $\beta$, and given
by (see formula~\ref{eq:sdiff})
\begin{equation}
f_{\pom/p}(\xpom,t) \sim \left( \frac{1}{\xpom} \right)^{2\apom(0)-1}
e^{2(b_0^{D}-\aprime \ln \xpom)t} \, .
\label{eq:pomflux}
\end{equation}
In this approach, the diffractive parton distributions would be
obtained as a convolution of the $\pom$ flux with parton distributions
in the $\pom$.

None of the approaches detailed above address the issue of the
dynamical origin of the $\pom$ exchange in perturbative QCD. The
mechanism for producing LRG is assumed to be present at some scale and
the evolution formalism allows to probe the underlying partonic
structure. A more dynamical approach will be discussed in the context of
the measurements performed up to date.

\section{Measurements of $\ftwod$ at HERA}

\begin{figure}[htb]
\begin{center}
\epsfig{file=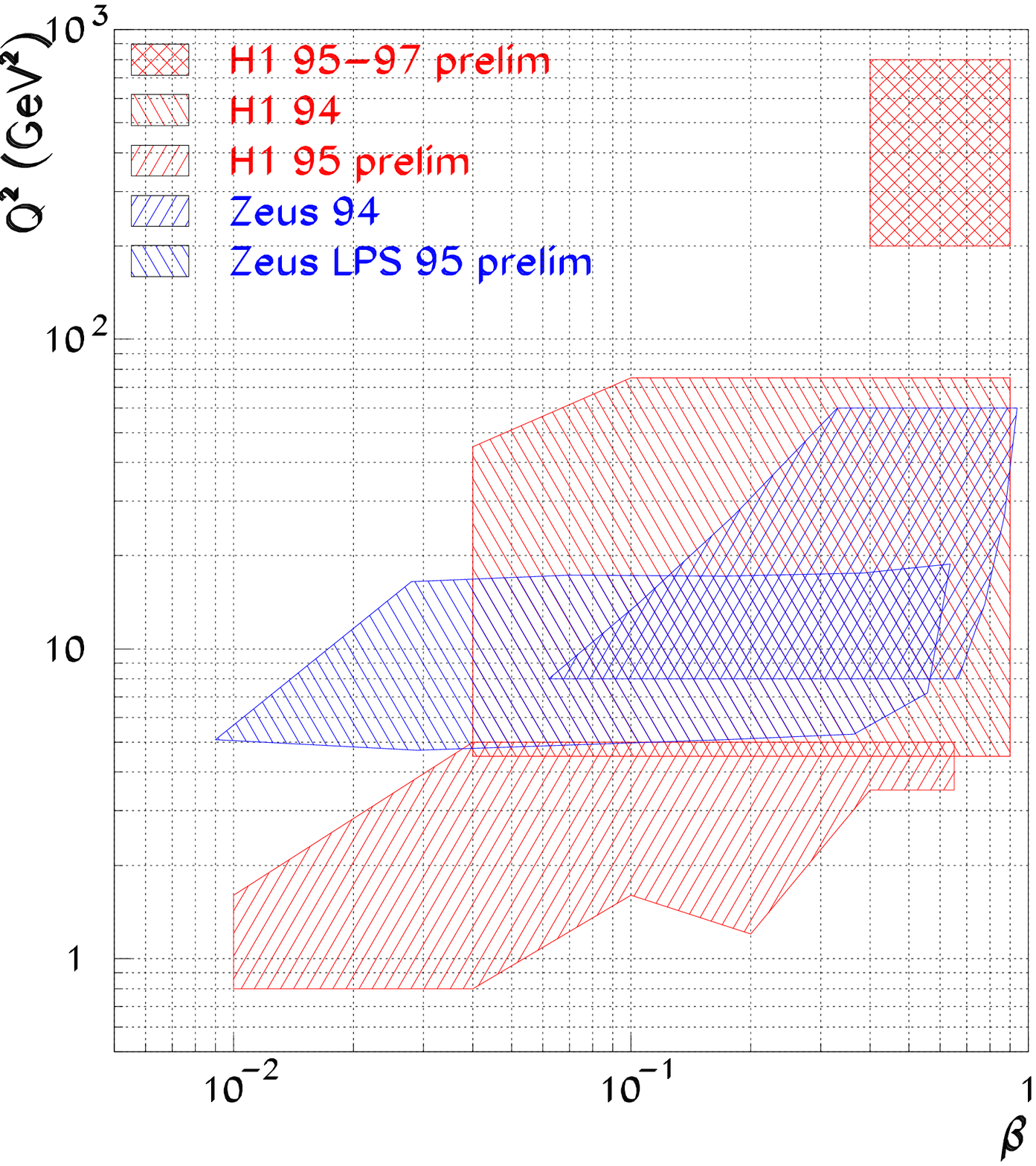,width=0.45\hsize} 
\epsfig{file=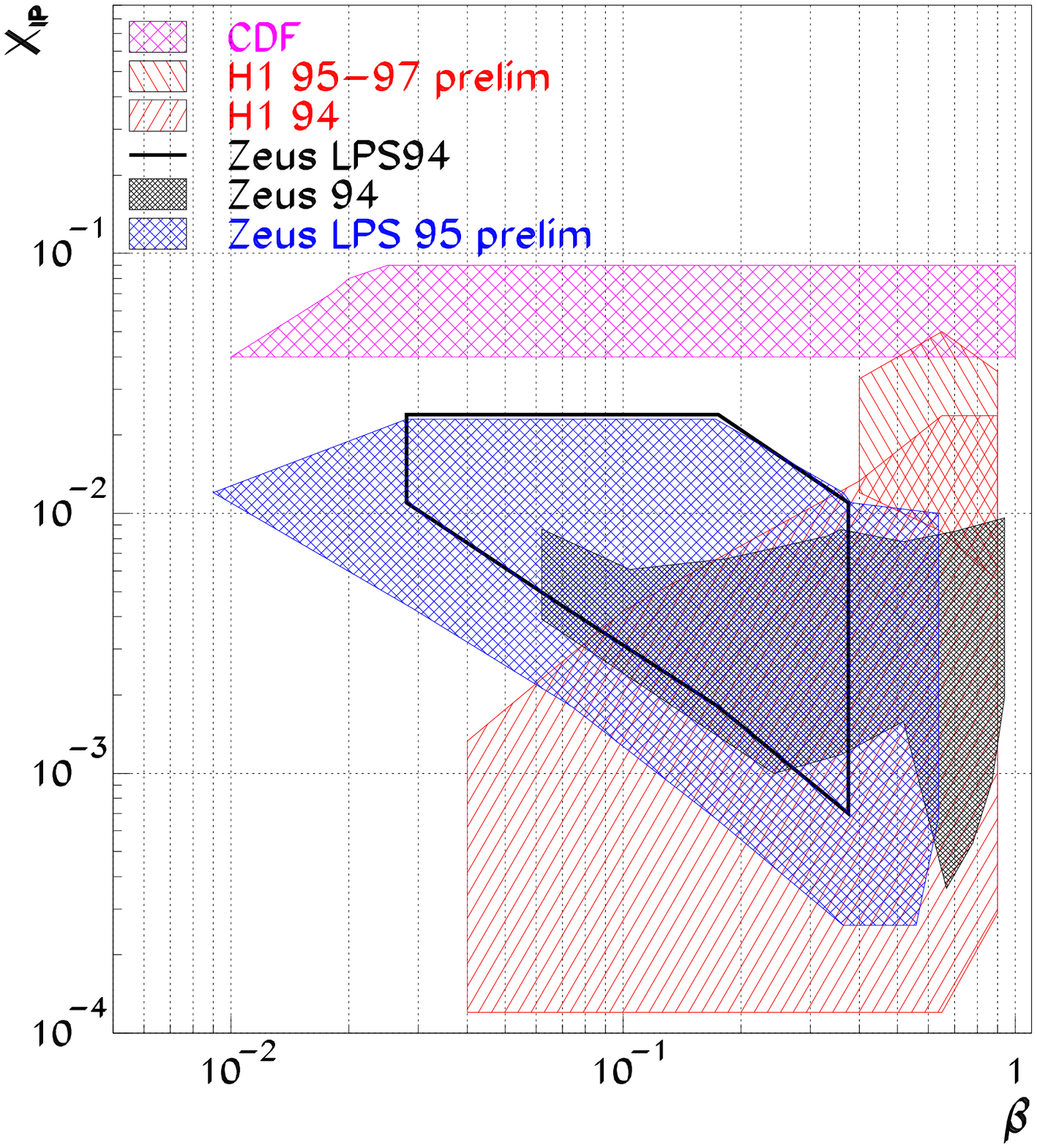,width=0.45\hsize} 
\caption{Phase space coverage in $\beta$, $Q^2$ and $\xpom$ for the 
  measurements of $\ftwod$ in $ep$ interactions. Also shown is the
  $\xpom$ and $\beta$ coverage of the recent CDF data with the recoil
  proton measured in the Roman pots (see section~\protect\ref{sec:cdfromanpots}).  }
\label{fig:contours} 
\end{center}
\end{figure}

The data analyzed to date come from $e^+p$ runs with the proton beam
momentum of 820 GeV and positron beam of 27.5 GeV. The coverage of
phase space in $\beta$ and $Q^2$ as well as $\beta$ and $\xpom$ is
shown in Figure~\ref{fig:contours}. 

\subsection{Energy dependence and Regge factorization \label{sec:energy}}

\begin{figure}[htb]
\begin{center}
\epsfig{file=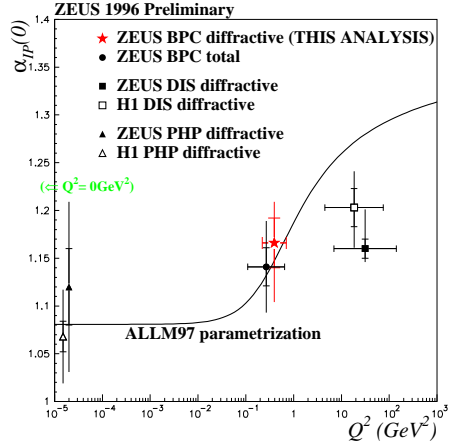,height=6cm} 
\caption{$Q^2$ dependence of $\apom(0)$ derived from measurements of 
  diffractive and total $\gvp$ cross sections. The curve (ALLM97)
  is a representation of the results obtained in
  inclusive DIS measurements.  }
\label{fig:apom0} 
\end{center}
\end{figure}

Following the lead of Regge phenomenology, the energy dependence of
$\sigma^D(\gvp)$ or the $\xpom$ dependence of $\ftwod$ in the data is
expressed in terms of $\apom(0)$. For $\xpom<0.01$ and within a given
range of $Q^2$ the data are compatible with a universal $\xpom$
dependence~\cite{h1-f2d3:1997,zeus-mx:1999,zeus-bpc-mx:1999}. The data
used for these studies are integrated over $t$. To derive the value of
$\apom(0)$, it is assumed that the slope of the $t$ distribution is
$Q^2$ independent and is given by $b^D=7
\gevmtwo$~\cite{zeus-lps:1998}.  For lack of other information
$\aprime=0.25 \gevmtwo$ is used. The $Q^2$ dependence of $\apom(0)$
derived from diffractive measurements is shown in
Figure~\ref{fig:apom0}. For comparison, also shown is the $Q^2$
dependence of $\apom(0)$ derived from the inclusive DIS measurements
and conveniently represented by the ALLM97
parameterization~\cite{allm:1997}.
As has been known for a while, the inclusive DIS data at high $W$ are
not compatible with a universal $\pom$ trajectory. They are however in
very good agreement with expectations of QCD evolution. The
diffractive measurements seem to point to some $Q^2$ dependence, but
less pronounced than that in the inclusive case. For $Q^2>10 \gevtwo$
the value of $\apom(0)$ is significantly higher than measured for soft
interactions.  The important observation is that the $W$ dependence of
diffractive DIS is the same as for the inclusive cross
section~\cite{zeus-mx:1999} and slower than would be expected from
Regge phenomenology.

\subsection{$Q^2$ and $\beta$ dependence}
\begin{figure}[htb]
\begin{center}
\setlength{\unitlength}{0.04\hsize}
\begin{picture}(15,11)
\put(0,0){\epsfig{file=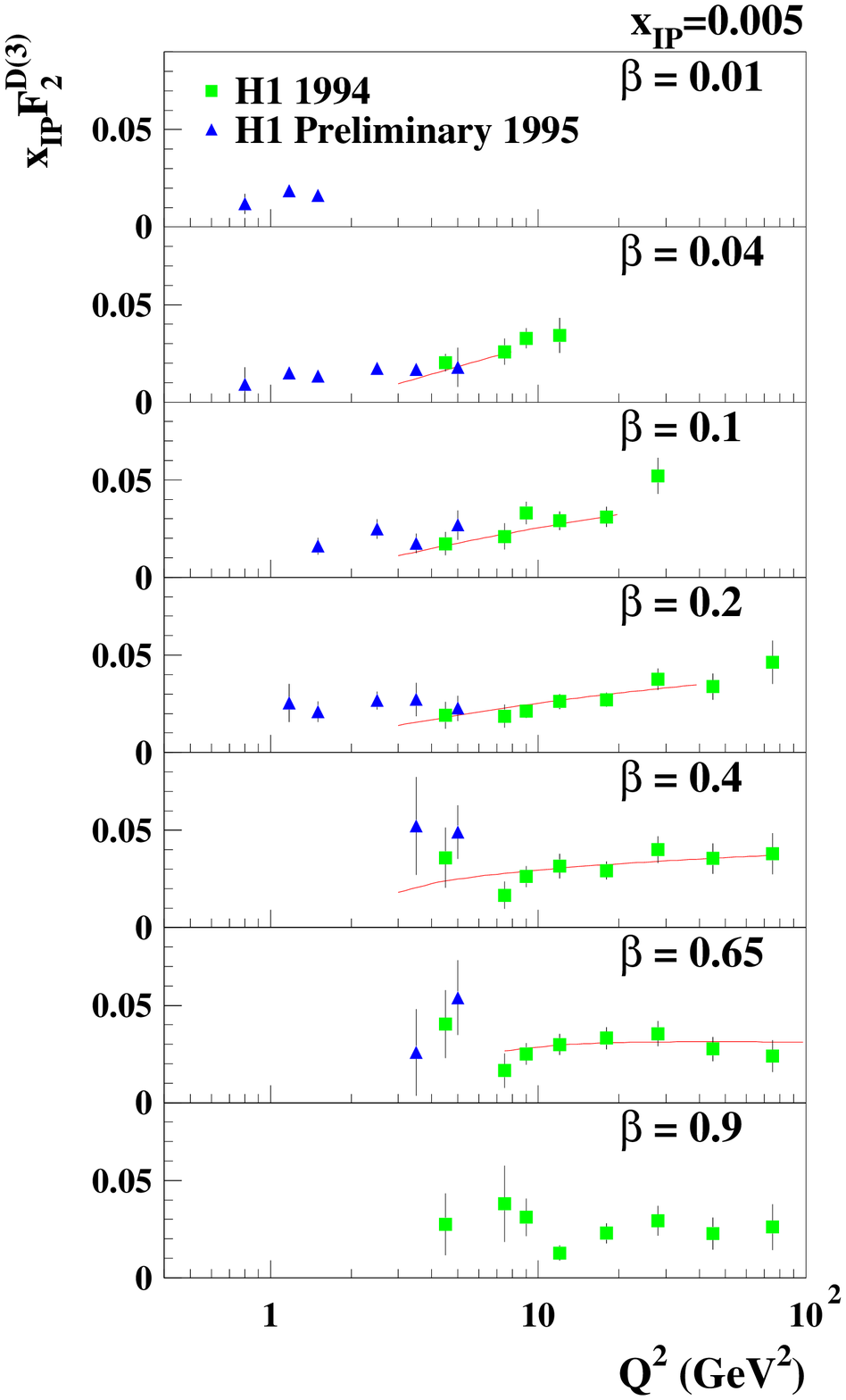, width=7\unitlength}}
\put(7.3,4.5){\epsfig{file=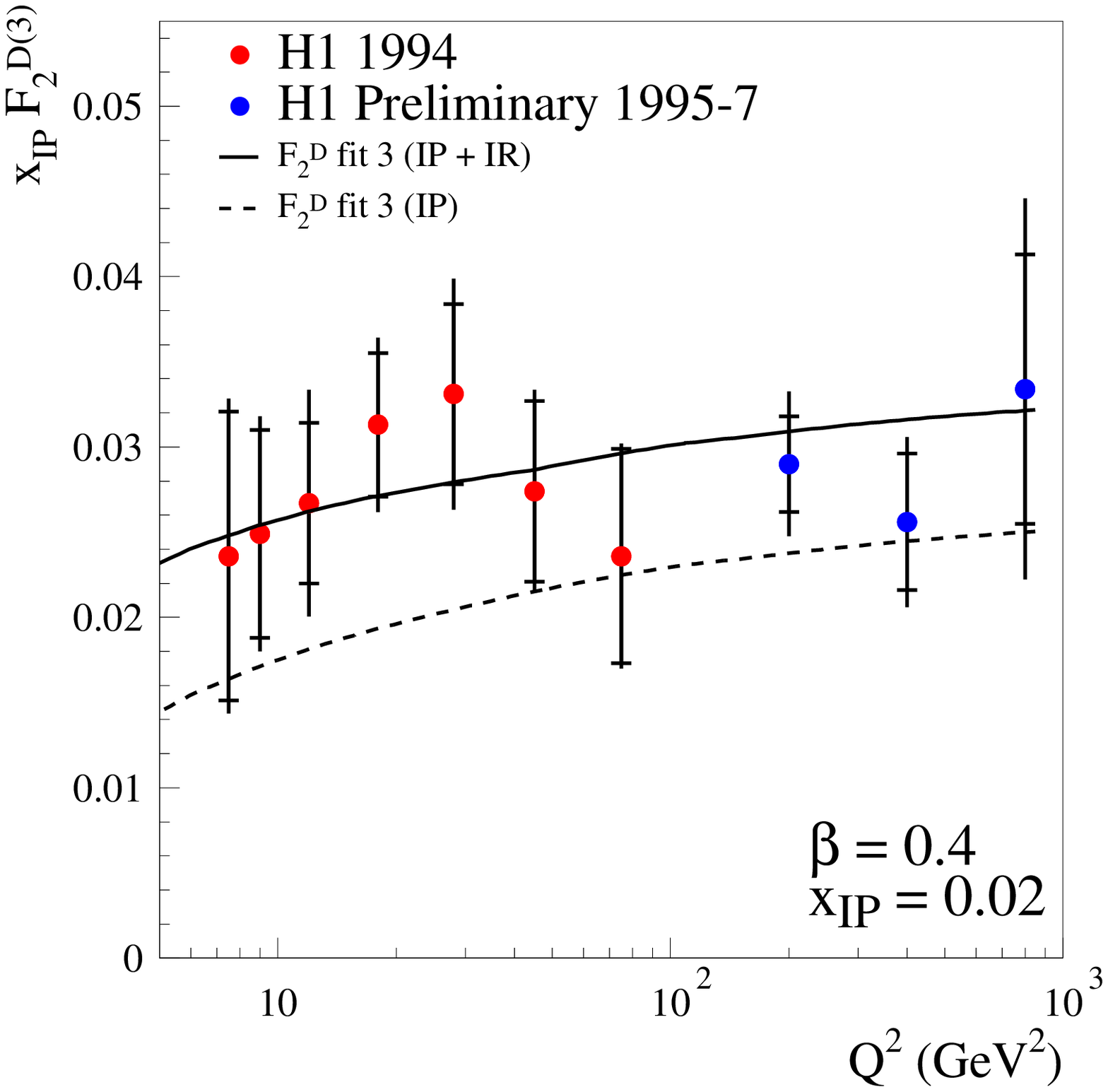,width=7\unitlength}}
\put(7.,0.2){\epsfig{file=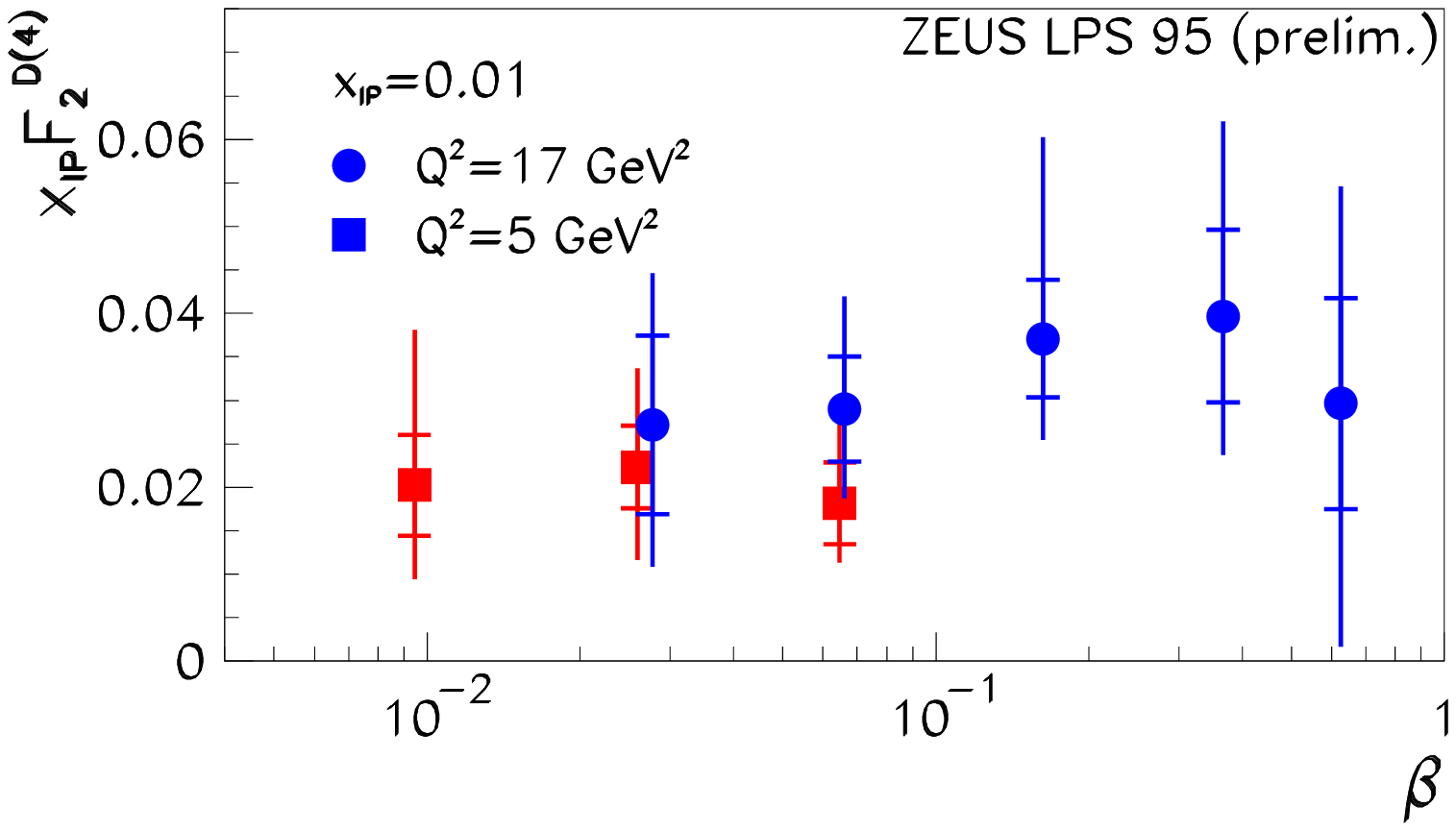,width=7\unitlength}}
\end{picture}
\vspace*{-0.5cm}
\end{center}
\caption{Left and upper right plots: $Q^2$ dependence of $\xpom \ftwodthree$ 
  at fixed values of $\beta$ and $\xpom$ as denoted in the figure.
  Lower right plot:$\beta$ dependence of $\xpom \ftwodfour$ (for
  $0.073<|t|<0.4 \gevtwo$) at fixed values of $Q^2$ and $\xpom$ as
  denoted in the figure.}
\label{fig:f2d4-q2}
\end{figure}

The $Q^2$ dependence of $\xpom\ftwodthree$ as measured by the H1
experiment~\cite{h1-f2d3:1997,h1-f2d3-prel} for two values
of $\xpom$ and for a range of $\beta$ values is shown in
Figure~\ref{fig:f2d4-q2}. Also shown in the figure is the $\beta$
distribution measured with the leading proton spectrometer (LPS) of
the ZEUS detector, at $\xpom=0.01$ for two values of
$Q^2$~\cite{zeus-lps-prel}.
Both the $Q^2$ and the $\beta$ dependence of $\ftwod$ are very
different from the one known for the proton $F_2$. There is almost no
$Q^2$ dependence for $\beta>0.4$ and the $\beta$ distribution itself
is relatively flat up to large $\beta \simeq 1$. This suggests a very
different parton content of diffractive scattering compared to the
inclusive scattering. As we will see below, for the DGLAP evolution to
hold, a large gluon component is required at large $\beta$ to
compensate for the loss of quark momenta in the evolution.

\subsection{Diffractive parton distributions \label{sec:partons}}
\begin{figure}[hbt]
\begin{center}
\setlength{\unitlength}{0.035\hsize}
\begin{picture}(16,9)
\put(-1.0,0){\epsfig{file=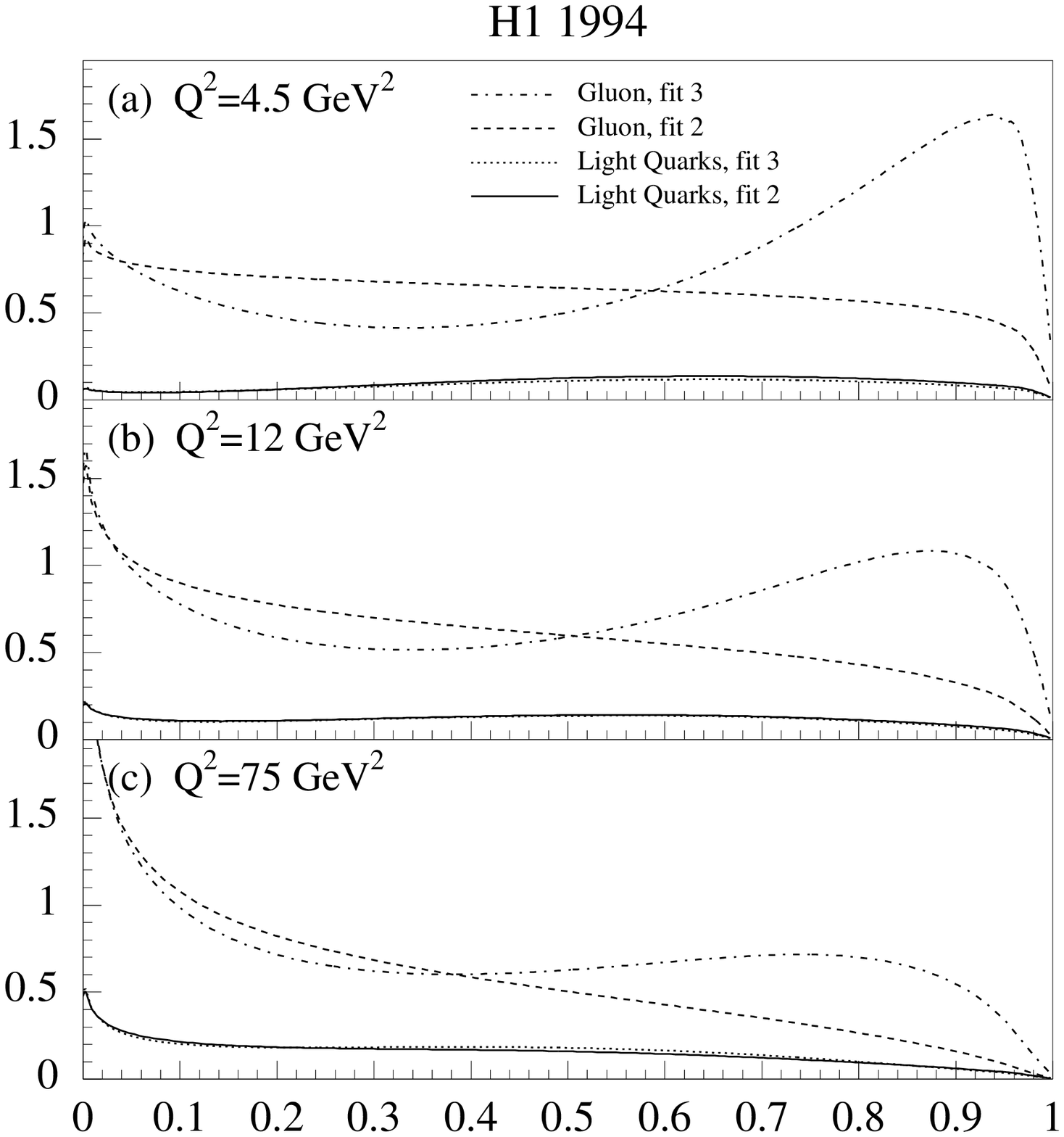,%
width=9\unitlength}}
\put(7.2,-0.6){$\beta$}
\put(-1.8,7){\rotatebox{90}{$\beta f(\beta)$}}
\put(9.,-0.8){\epsfig{file=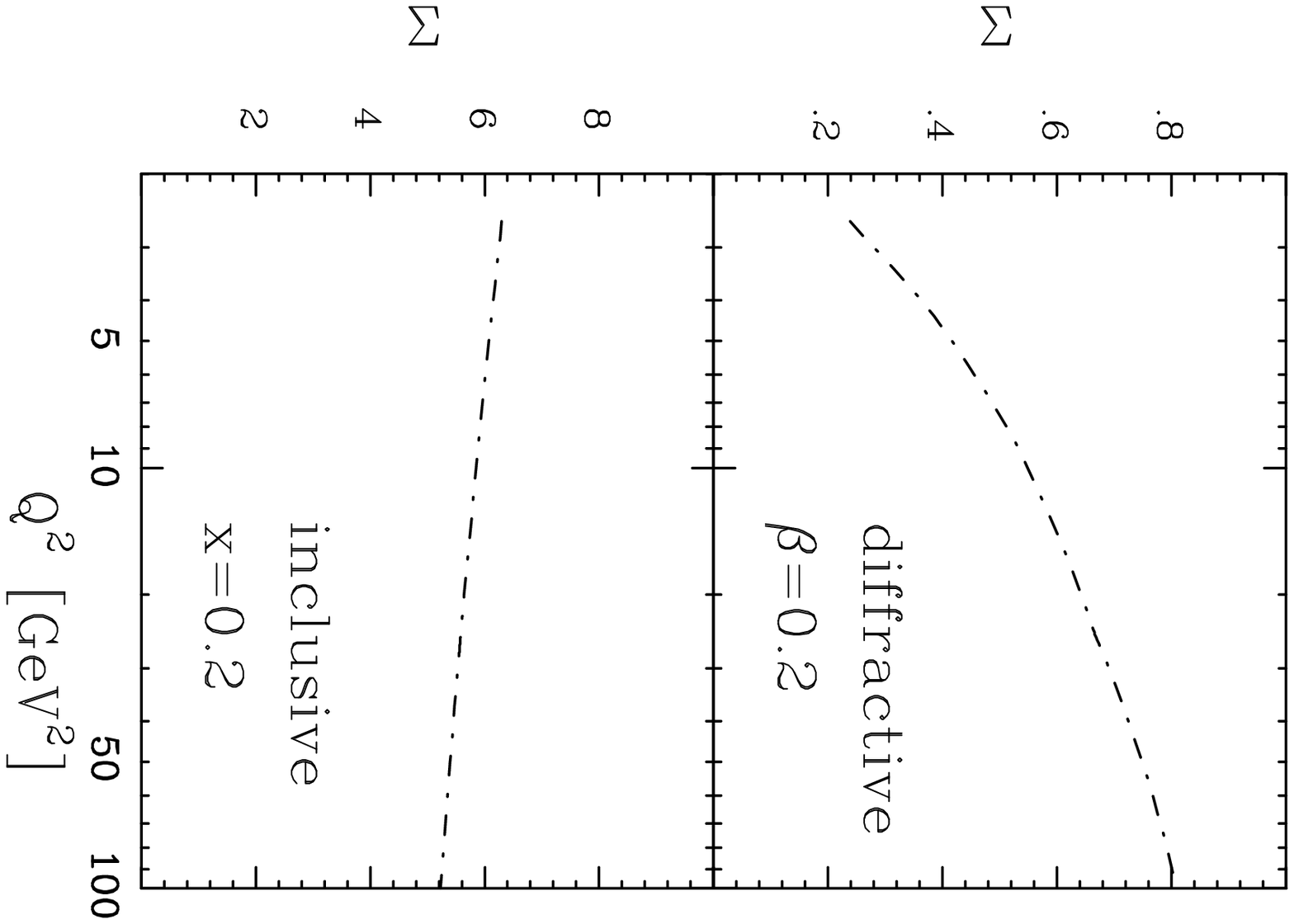,%
width=10\unitlength,angle=90}}
\end{picture}
\end{center}
\vspace*{-0.3cm}
\caption{Left: diffractive parton densities $\beta f(\beta )$ as 
  a function of $\beta$ obtained by fitting the DGLAP evolution
  equations.  Right: difference in $Q^2$ evolution of the singlet
  structure function $\Sigma$ for diffractive and inclusive parton
  distributions for $\beta=0.2$ and $x=0.2$ respectively.}
\label{fig:partons}
\vspace*{-0.5cm}
\end{figure}

The diffractive parton density functions \index{diffractive parton
  densities} (DPDF) are derived from $\ftwod$ by fitting the DGLAP
evolution equations with postulated input $\beta$ distributions at
some low $Q^2_0$ scale. The fit consists of adjusting the parameters
of the input distributions to get the best description of all
available data. No momentum sum-rule is available for diffractive
partons, however one assumes flavor symmetry for light quarks.  The
results of such a fitting procedure~\cite{h1-f2d3:1997} are shown in
Figure~\ref{fig:partons}.
Different solutions are found, however for all cases a large gluon
component is required. The net result is that while the density of
inclusive quark distributions (represented by the flavor singlet
structure function $\Sigma$) decreases with $Q^2$ at $x=0.2$, the
density of diffractive partons is still rising with $Q^2$ at the
equivalent $\beta=0.2$~\cite{Hautmann:1999}.

\subsection{Jet and charm production}
\begin{figure}[bht]
\begin{center}
\hspace*{-2cm}\epsfig{file=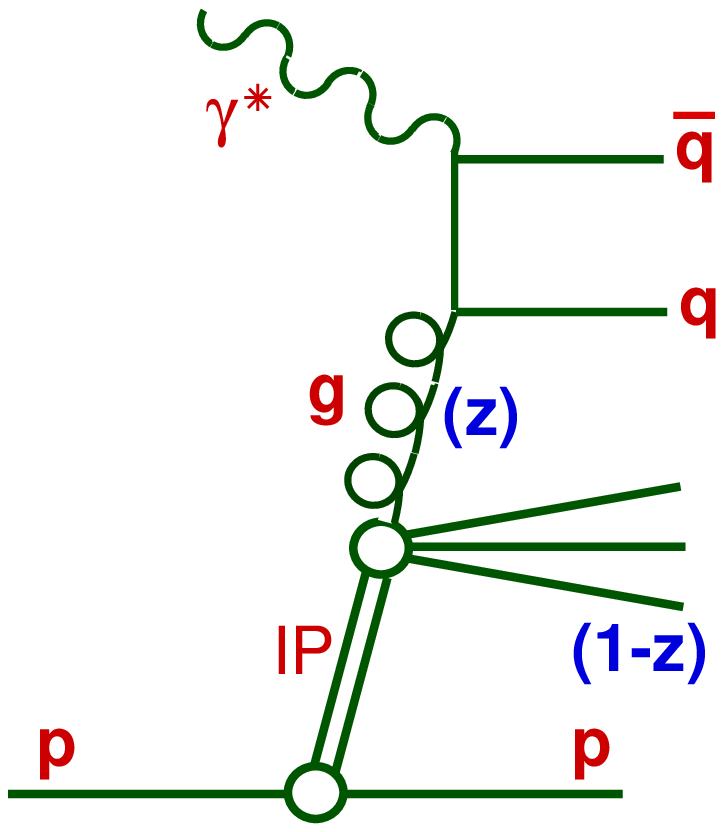,height=4.5cm}
\hspace*{1cm}\epsfig{file=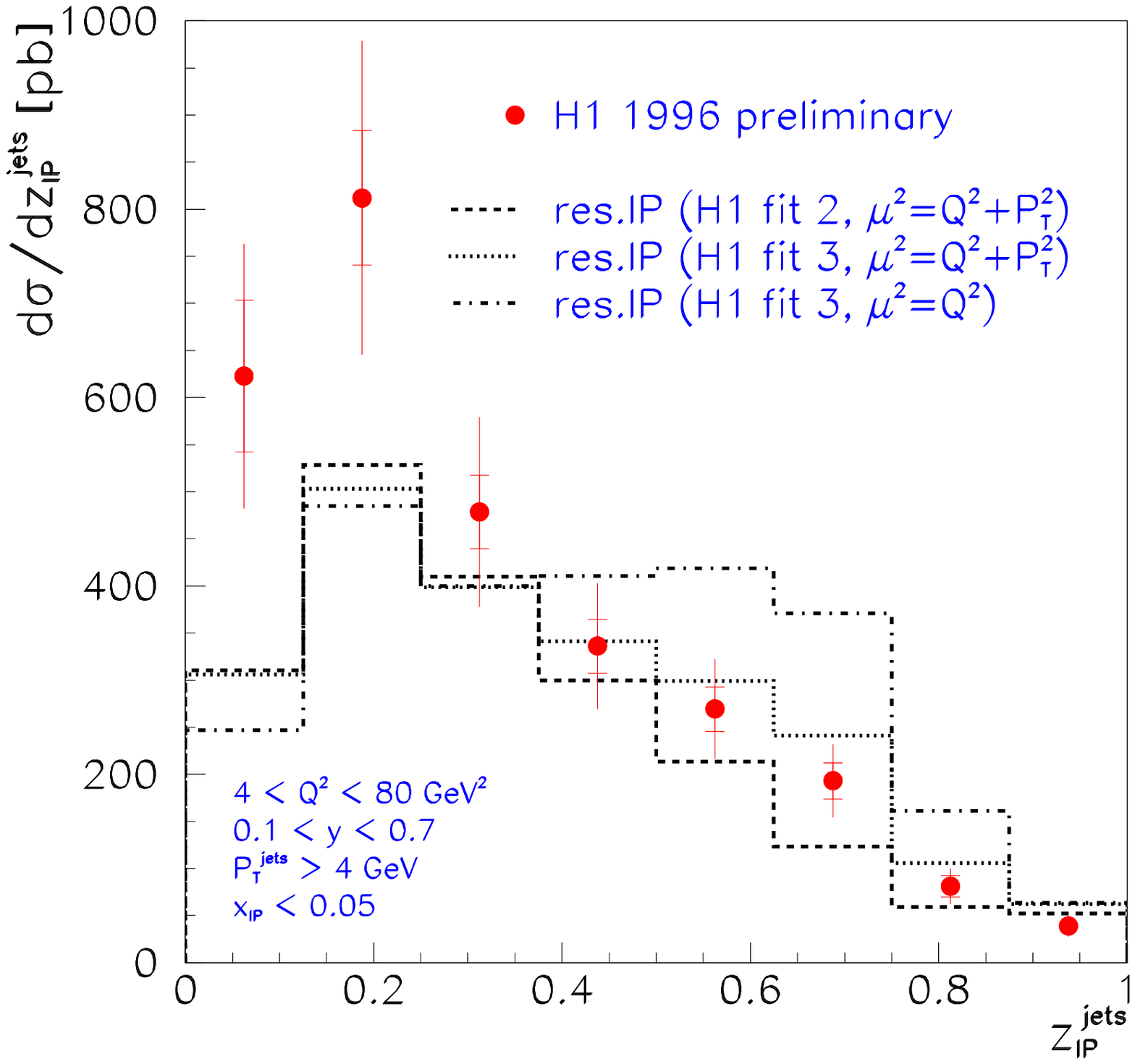,height=5cm} 
\caption{Left: boson-gluon fusion diagram. Right: The dependence 
  of diffractive dijet cross section on $z^{\mathrm jets}_{\pom}$, the
  experimental estimate of $z$ as defined in the left diagram. The
  curves correspond to expectations for different DPDFs. }
\vspace*{-0.5cm}
\label{fig:h1-jets} 
\end{center}
\end{figure}

The partonic structure of diffractive scattering can be further
explored by studying the hadronic final states. A large gluon component
is expected to lead to a copious production of high $p_T$ jets as well
as of charm, through the boson-gluon fusion diagram depicted in
Figure~\ref{fig:h1-jets}. 
The H1 experiment has measured the cross section for dijet production
in diffractive scattering~\cite{h1-dijets}. The observed rates are in
agreement with expectations based on DPDFs obtained from fits to
$\ftwod$.

Diffractive charm production has also been
measured~\cite{h1-charm,zeus-charm}, however while ZEUS measurements
are in agreement with expectations, the H1 result seems to be smaller
than expected. Note that the statistical and systematic errors of
these measurements are quite large and a definite answer awaits more
data.

\section{Generic model for diffractive DIS \label{sec:ddis}}
A tremendous theoretical effort has been, and still is, devoted to the
understanding of the dynamics behind diffractive DIS (for a recent
review see~\cite{RMP,Hebecker:1999} and references therein\footnote{
  List of representative papers from which a thorough exploration can
  begin:~\protect\cite{Bartels:1999,Bialas:1998,Buchmuller:1999,Golec-Biernat:1999,Gotsman:1997,Hautmann:1998,Nikolaev:1997,Ryskin:1996}.}).
A simple picture of diffraction emerges if the process is viewed in
the rest frame of the proton. The virtual photon develops a partonic
fluctuations, whose lifetime is $\tau
=1/2m_px$~\cite{Ioffe:1984}. At the small $x$ typical of HERA, where $\tau
\sim 10 - 100 \units{fm}$, it is the partonic state rather than the
photon that scatters off the proton. If the scattering is elastic, the
final state will have the features of diffraction.

\begin{figure}[htb]
\begin{center}
\epsfig{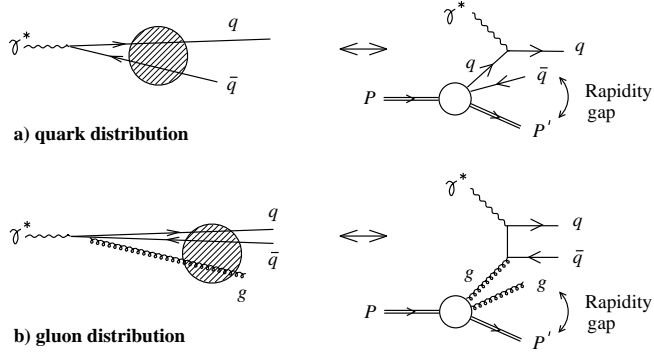} 
\caption{Schematic representation of the equivalence between diffractive 
  DIS in the proton rest frame (left) and in the Breit frame (right).  }
\label{fig:equivalence} 
\end{center}
\end{figure}

The equivalence between the approach in the proton rest frame and in
the Breit frame is schematically depicted in
Figure~\ref{fig:equivalence}~\cite{Buchmuller:1999a}. An interaction
initiated by fluctuation of $\gv$ into a $q\bar{q}$ ($q\bar{q}g$)
state will contribute to the diffractive quark (gluon) distribution.
The fluctuations of the $\gv$ are described by the wave functions of
the transversely and longitudinally polarized $\gv$ which are known
from perturbative QCD. Small and large partonic configurations of the
photon fluctuation are present. For large configurations
non-perturbative effects dominate in the interaction and the
treatment of this contribution is subject to modeling.  For a small
configuration of partons (large relative $k_T$) the total
interaction cross section of the created color dipole on a proton target is
given by~\cite{Blaettel:1993,Frankfurt:1999}
\begin{eqnarray}
\sigma_{q\bar{q}p}&=&\frac{\pi^2}{3}r^2\alpha_S(\mu)xg(x,\mu) \, ,
\label{eq:qqp} \\
\sigma_{q\bar{q}gp}&\simeq& \sigma_{ggp}=\frac{9}{4}\sigma_{q\bar{q}p} \, ,
\label{eq:qqgp}
\end{eqnarray}
where $r$ is the transverse size of the color dipole and $\mu \sim
1/r^2$ is the scale at which the gluon distribution $g$ of the proton
is probed. The corresponding elastic cross section is obtained from
the optical theorem. In this picture, the gluon dominance in
diffraction results from the dynamics of perturbative QCD (see
equation~(\ref{eq:qqgp})).

The interesting observation is that all the models, whether starting
from a purely perturbative approach~\cite{Hautmann:1999} or a
semi-classical approach~\cite{Buchmuller:1999}, give a good
description of the $Q^2$ and $\beta$ dependence of $\ftwod$. The
leading-twist behavior of diffractive DIS is determined by the
hadronic nature of the $\gvp$ interactions. The $\beta$ dependence
seems to be mostly determined by the wave function of $\gv$.

A possible explanation of the success of the perturbative approach to
diffractive DIS has been proposed by Mueller~\cite{Mueller:1998}.  If
the large size $\gv$ fluctuations were to be absorbed, diffraction
would be dominated by small size configurations which lend themselves
to perturbative calculations. The absorption would be a manifestation
of unitarity effects \index{unitarity effects} and would explain the
slower than expected $W$ dependence of the $\gvp$ diffractive cross
section (see section~\ref{sec:energy}). Absorption effects have been
successfully incorporated in some of the
models~\cite{Golec-Biernat:1999,Gotsman:1997}.

The possibility that unitarity effects at low $x$ are not negligible
has been pointed out by Frankfurt and Strikman~\cite{Frankfurt:1999}.
The probability that an interaction on parton $i$ is associated with a
diffractive final state, $P^D_i$, is defined as
\begin{equation}
P^D_i(x,Q^2)=\frac{\int_{t,\xpom} f^D_i\left(\frac{x}{\xpom},Q^2,
\xpom,t\right)d\xpom dt}{f_i(x,Q^2)} \, ,
\label{eq:pdiff}
\end{equation}
where $f_i$ stands for the inclusive density of parton $i$.  $P^D_i$ may
be interpreted as the ratio of elastic to total cross sections,
$r_{el}(i)=\sigma_{el}(i)/\sigma_{tot}(i)$, induced either by a
$q\bar{q}$ ($i=q$) or $q\bar{q}g$ ($i=g$) configurations of
$\gv$. In the black disc limit, $r_{el} = 0.5$. In DIS, for $Q^2=4.5
\gevtwo$ and $x<10^{-3}$, the calculation using the
ACTW~\cite{Alvero:1999} parameterization leads to $P^D_g\simeq 0.4$
for gluons and $P^D_q \simeq 0.15$ for quarks. The value of $P^D_g$,
close to the black disc limit, may indicate the presence of unitarity
effects in the gluon sector of the evolution.

\section{Diffraction in $p\bar{p}$ scattering}

\begin{figure}[htb]
\begin{center}
\setlength{\unitlength}{0.0035\hsize}
\begin{picture}(90,80)
\put(0,0){\epsfig{file=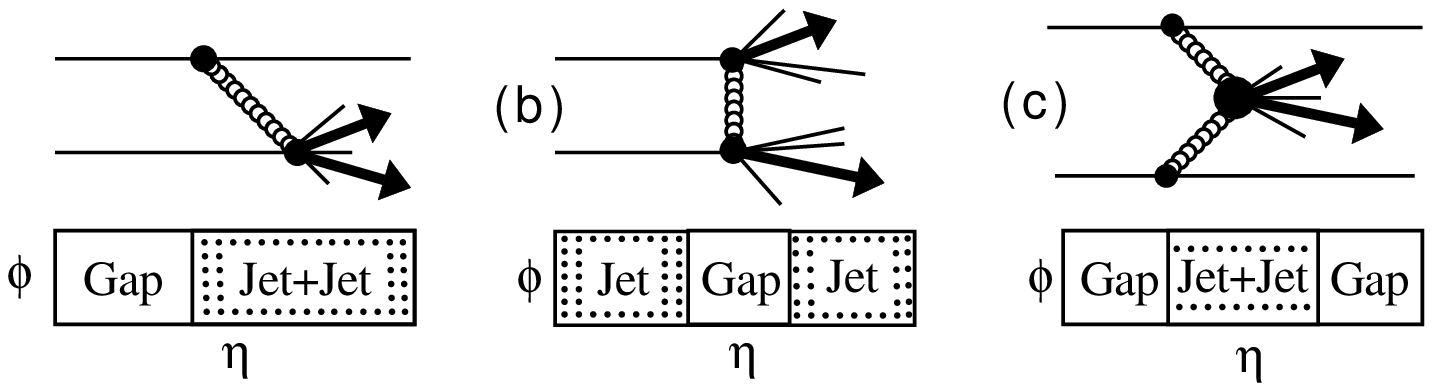,%
bbllx=75,%
bblly=636,%
bburx=197,%
bbury=733,%
clip=,%
width=80\unitlength}}
\put(2,35){\scalebox{1.2}{$p,\bar{p}$}}
\put(2,63){\scalebox{1.2}{$\bar{p},p$}}
\end{picture}
\end{center}
\caption{Schematic representation of single diffractive events in 
  $p\bar{p}$ collisions and the corresponding density of the phase
  space of the azimuthal angle $\phi$ and of pseudorapidity $\eta$.
  }
\label{fig:fnal-diag}
\end{figure}

Diffractive-like events have been observed in hard $p\bar{p}$
scattering by both the CDF and the D0 experiments.  The signature of these
events, depicted in Figure~\ref{fig:fnal-diag}, is the presence of a
large rapidity gap in one of the forward regions associated with the
presence of either large transverse momentum
jets~\cite{cdf-jets,d0-jets}, beauty~\cite{cdf-beauty},
$J/\psi$~\cite{cdf-jpsi} or the $W^{\pm}$ boson~\cite{cdf-w}. 

\subsection{Rates of diffractive processes}

\begin{figure}[bht]
\begin{center}
\epsfig{file=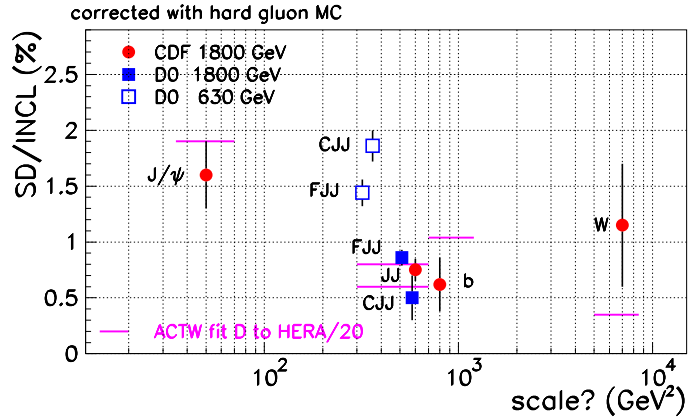,width=0.6\hsize}
\end{center}
\vspace*{-0.5cm}
\caption{Ratio of single 
  diffraction to inclusive production in $p\bar{p}$, SD/INCL, as a
  function of the approximate hard scale in the process. JJ stands for two
  jet production while C and F stand for forward and central
  production.  Horizontal bars are expectations obtained using the
  ACTW parameterization, fit D~\protect\cite{Alvero:1999}, scaled down
  by factor 20.  }
\label{fig:fnal-comp}
\end{figure}

A compilation of the published and preliminary results on the ratio of
diffractive to inclusive events with a given hadronic final state is
presented in Figure~\ref{fig:fnal-comp} as a function of the typical
average scale of the process.  The measurements are compared to
expectations obtained using the ACTW-fit D
parameterization~\cite{Alvero:1999} of DPDFs. In order to reproduce
vaguely the measured numbers, the expectations have to be scaled down
by a factor 20. This may not be surprising as QCD factorization is not
applicable to hard diffractive scattering in hadron
collisions~\cite{Collins:1998}. What might be surprising is that the
overall scale dependence is quite well reproduced.  Unfortunately
there are no calculations available for $p\bar{p}$ collisions at
$\sqrt{s}=630 \gev$.  The difference in the contribution of the
diffractive quarks and gluons to different processes has been used by
CDF to constrain, to leading order, the momentum fraction $f_g$ of
gluons relative to all diffractive partons.  Including $W$, jet and
beauty production, $f_g=0.54^{+0.16}_{-0.14}$, marginally less than in
diffractive DIS, where this ratio is close to $90\%$, for
next-to-leading order partons.

\subsection{$\xi$ and $\beta$ dependence \label{sec:cdfromanpots}} 

The fractional momentum loss $\xi$ of the proton  in hard diffractive
jet production has been measured by D0~\cite{d0-jets}. For
$\sqrt{s}=1800~(630)\gev$, the distribution peaks at $\xi\simeq
0.03~(0.07)$ and extends to $\xi=0.1~(0.2)$. This is a region where
contributions from sub-leading trajectories may become important. 

\begin{figure}[htb]
\begin{center}
\epsfig{file=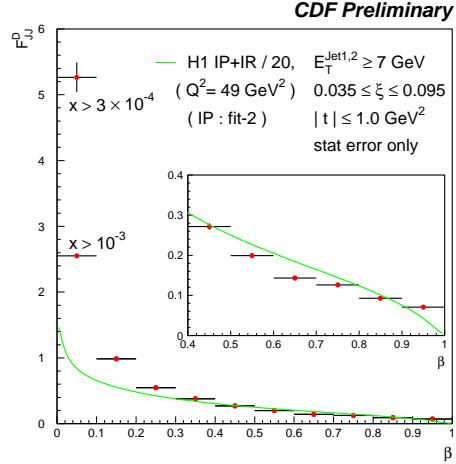,width=0.4\hsize}
\end{center}
\vspace*{-0.5cm}
\caption{$\beta$ distribution of two jet production with a leading 
  $\bar{p}$ in $p\bar{p}$ collisions compared to expectations based on
  diffractive parton distributions obtained at
  HERA~\cite{h1-f2d3:1997} scaled down by factor 20.  }
\label{fig:roman-pots}
\end{figure}

CDF has measured $\xi$ and $\beta$ by detecting the leading $\bar{p}$
in a forward spectrometer consisting of Roman
pots~\cite{cdf-romanpots}. The results are expressed in terms of an
effective structure function $F^D_{JJ}$ derived from the rate of
leading $\bar{p}$ and the two-jet inclusive cross section. The shape
of the $\beta$ distribution was found to be independent of $\xi$,
suggesting Regge factorization.  For the measured range,
$0.04<\xi<0.09$, the $\xi$ dependence of $F^D_{JJ}$ at $\beta=0.2$ is
described by $\xi^{-0.87}$. The $\beta$ distribution of $F^D_{JJ}$ is
shown in Figure~\ref{fig:roman-pots} and compared to expectations
using the H1 - fit 2 parameterization~\cite{h1-f2d3:1997}.  To get the
right order of magnitude for $\beta>0.4$, the expectations have to be
scaled down by a factor 20. In addition, CDF observes a relatively
higher yield of low $\beta$ events.
Without the power of the QCD factorization theorem and given the very
different $\xi$ range, it is very difficult to draw any conclusions
from the difference in the $p\bar{p}$ and $ep$ data.

\section{Vector meson production}
\index{vector meson production}
\begin{figure}[htb]
\begin{center}
\epsfig{file=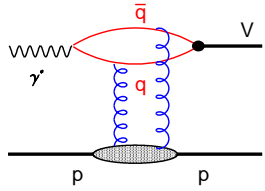,width=0.25\hsize}
\end{center}
\vspace*{-0.5cm}
\caption{Schematic diagram for exclusive vector meson, $V$, production 
in hard diffractive scattering.  }
\label{fig:vm-diag}
\end{figure}
To test ideas behind the generic model of diffractive DIS (see
section~\ref{sec:ddis}) and its hard component, one needs to devise a
trigger to isolate when the photon fluctuates into a small-size
$q\bar{q}$ configuration. A small-size $q\bar{q}$ dipole is most
likely to be produced if the virtual photon is longitudinally
polarized or if the dipole consists of heavy quarks such as $c\bar{c}$
or $b\bar{b}$.  The exclusive $J/\psi$~\cite{Ryskin:1993} and
$\Upsilon$ photoproduction as well as vector meson production in
DIS~\cite{Brodsky:1994} are likely candidates for hard diffractive
processes fully calculable in perturbative QCD. The corresponding
diagram is shown in Figure~\ref{fig:vm-diag}.
The predictions of perturbative QCD are very distinct:\\
{\bf (1)} The flavor dependence of $\sigma_V=\sigma (\gvp \rightarrow
Vp)$ is determined by the photon quark current and the SU(4) relation
\begin{equation}
\sigma_{\rho}:\sigma_{\omega}:\sigma_{\Phi}:\sigma_{J/\psi} = 9:1:2:8 \, ,
\label{eq:su4}
\end{equation}
which is badly broken in soft interactions (see
eg.~\cite{Bauer:1978}), is restored.\\
{\bf (2)} Exclusive $V$ production in DIS is dominated by the
longitudinal $\gamma^*_L$ component and~\cite{Brodsky:1994}
\begin{equation}
\frac{d\sigma_{V,L}}{dt}=\frac{A}{Q^6}\alpha_S(Q^2)|xg(x,Q^2)|^2 \, ,
\label{eq:sigmal}
\end{equation}
where $A$ depends on the $V$ wave function. At low $x$, the gluon
distribution increases rapidly when $x$ decreases ($W$ increases) and
therefore a rapid increase of $\sigma_V$ with $W$ is expected.\\
{\bf (3)} Because of the fast increase with $Q^2$ of gluons at low
$x$, the expected $Q^2$
dependence is slower than $Q^{-6}$.\\
{\bf (4)} The $t$ distribution should be universal, determined by the
two-gluon form-factor, with $\aprime \simeq 1/Q^2$, because of the
perturbative nature of the interaction.

Below some of the recent experimental findings are reviewed. A
detailed discussion of HERA results can be found in~\cite{RMP}.

\subsection{Flavour dependence}
\begin{figure}[htb]
\begin{center}
\epsfig{file=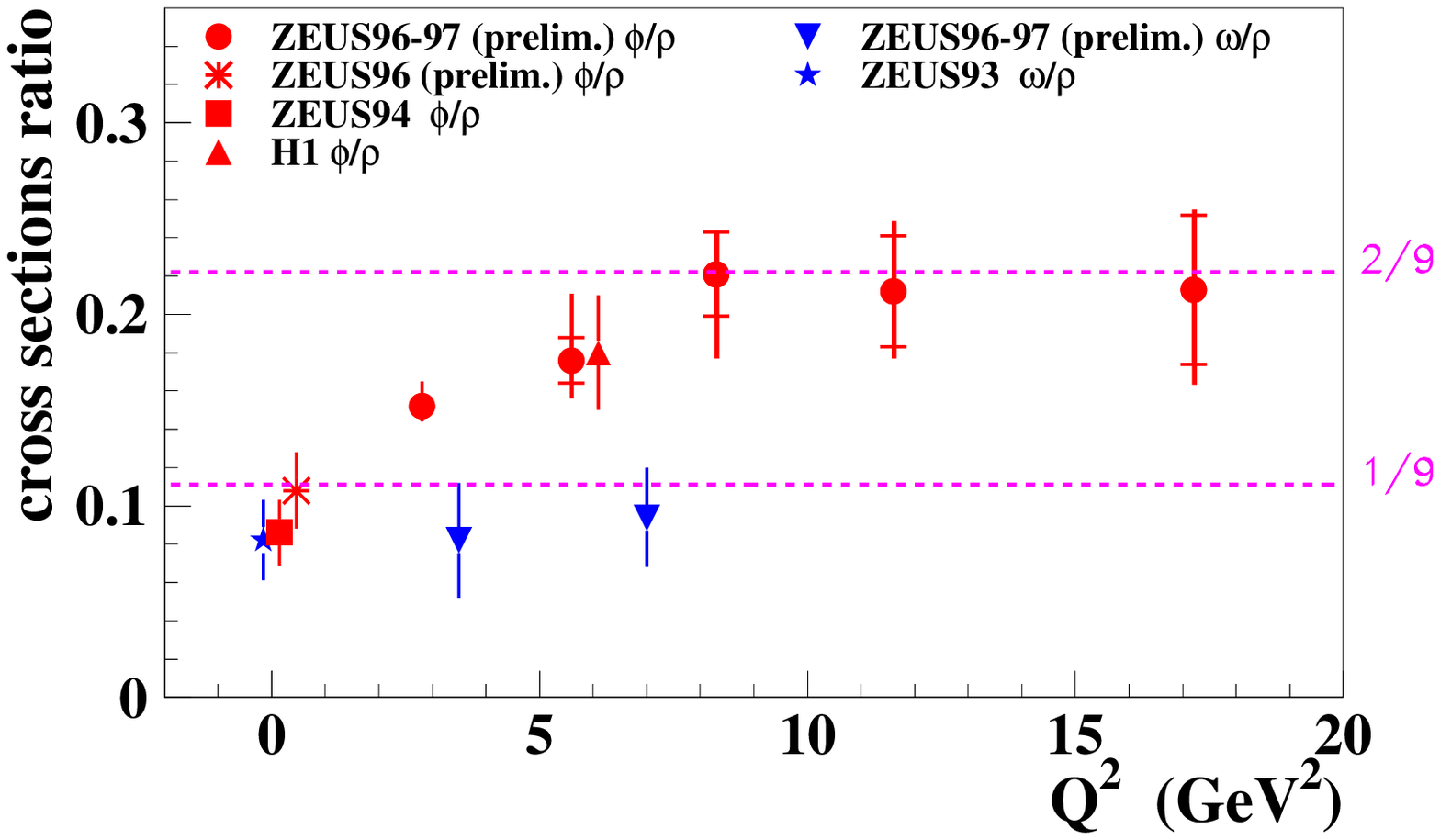,width=0.4\hsize}
\epsfig{file=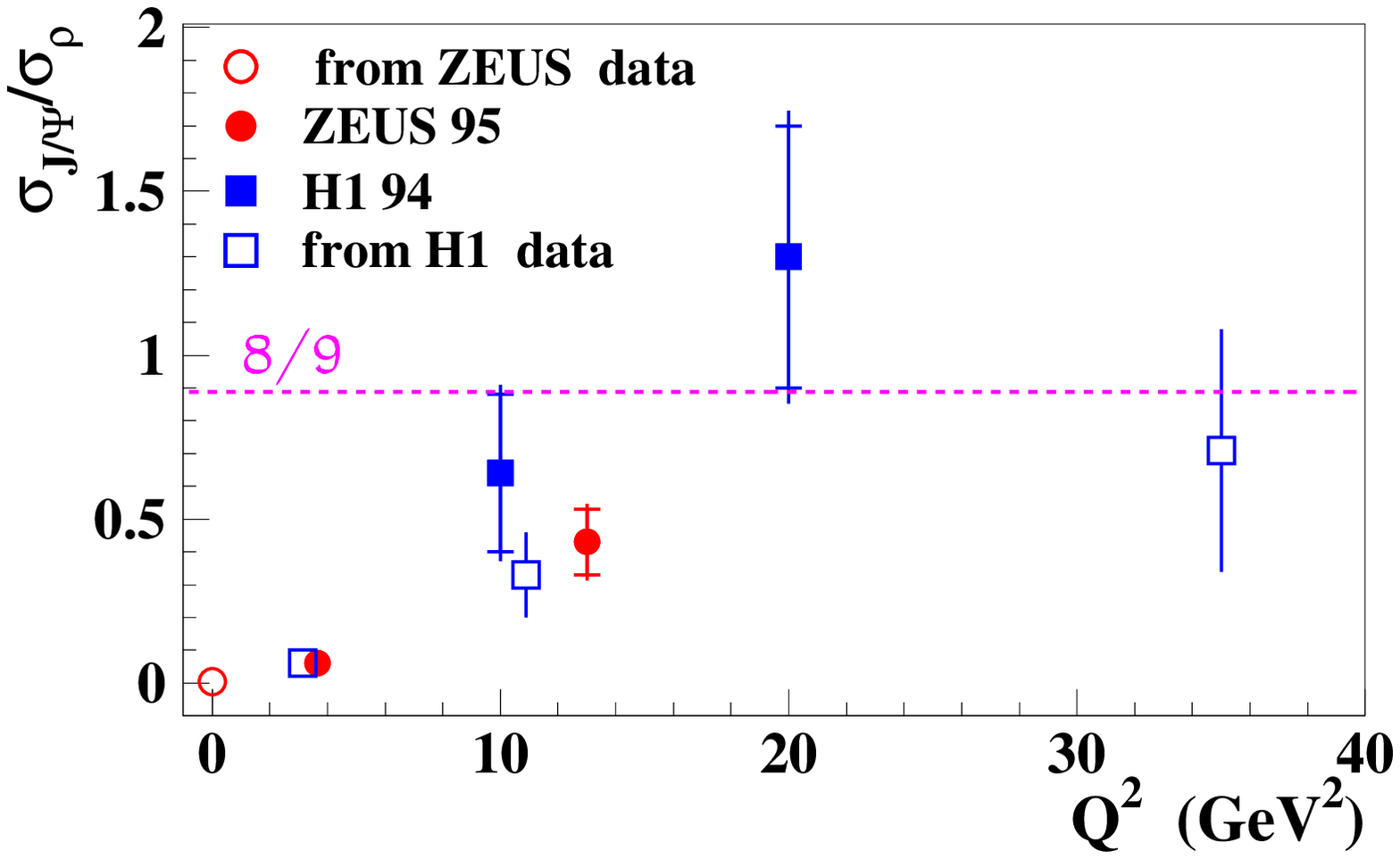,width=0.4\hsize}
\end{center}
\vspace*{-0.5cm}
\caption{Cross sections ratio $\sigma_{V_1}/\sigma_{\rho^0}$ for 
  $V_1=\Phi$, $\omega$ (left) and $J/\psi$ (right) as a function of
  $Q^2$.  }
\label{fig:su4}
\end{figure}

A compilation of the ratio of $\Phi$, $\omega$ and $J/\psi$ production
cross sections to that of the $\rho^0$ as a function of $Q^2$ is shown
in Figure~\ref{fig:su4}.
The $\Phi:\rho^0$ and $J/\psi:\rho^0$ ratios increase with $Q^2$ and
at high $Q^2$ reach, within errors, the value expected from SU(4). The
$\omega:\rho^0$ remains constant.

\subsection{$W$ dependence}

\begin{figure}[hbt]
\begin{center}
\epsfig{file=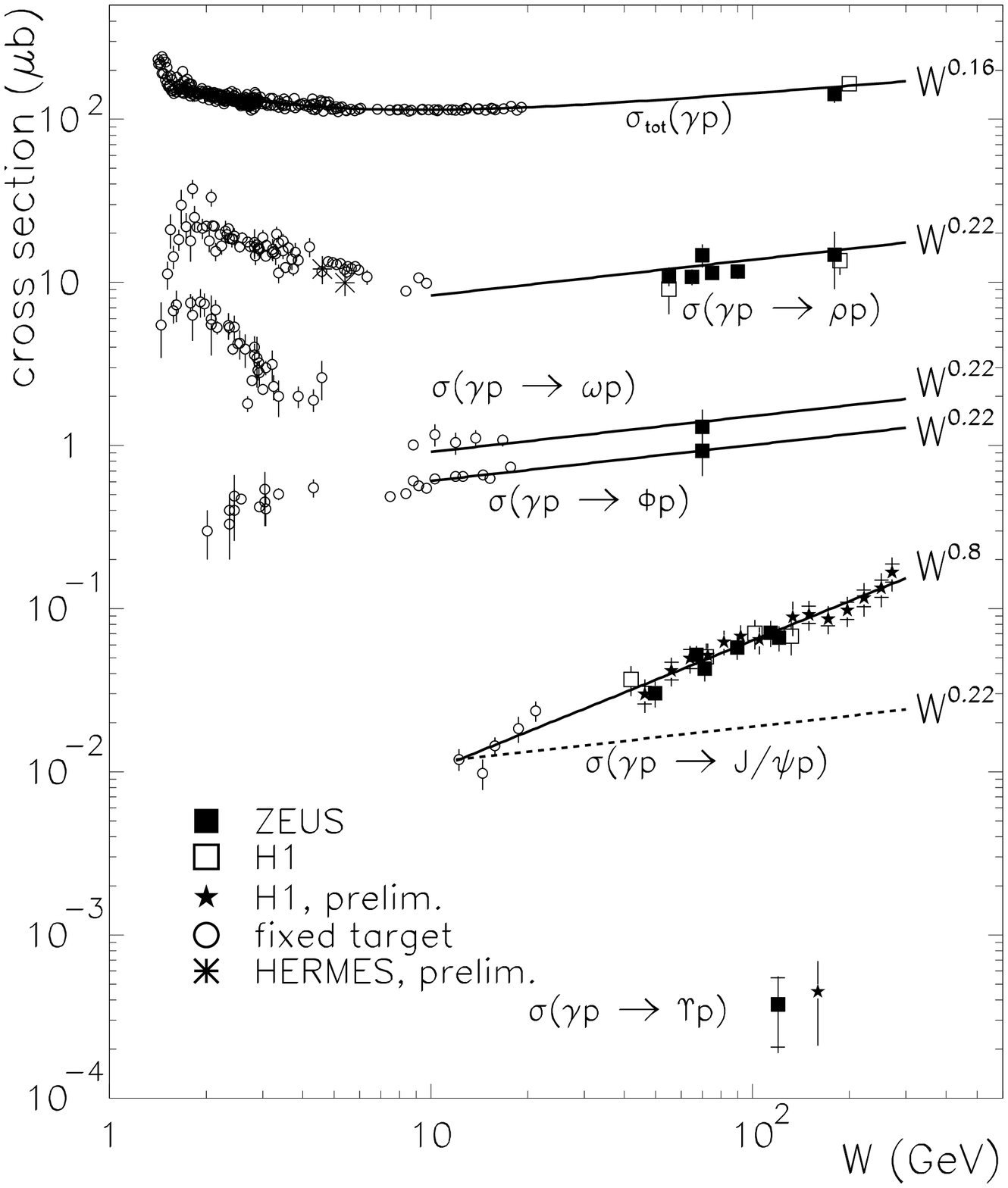,width=0.40\hsize,%
bbllx=17,bblly=98,bburx=534,bbury=708}
\epsfig{file=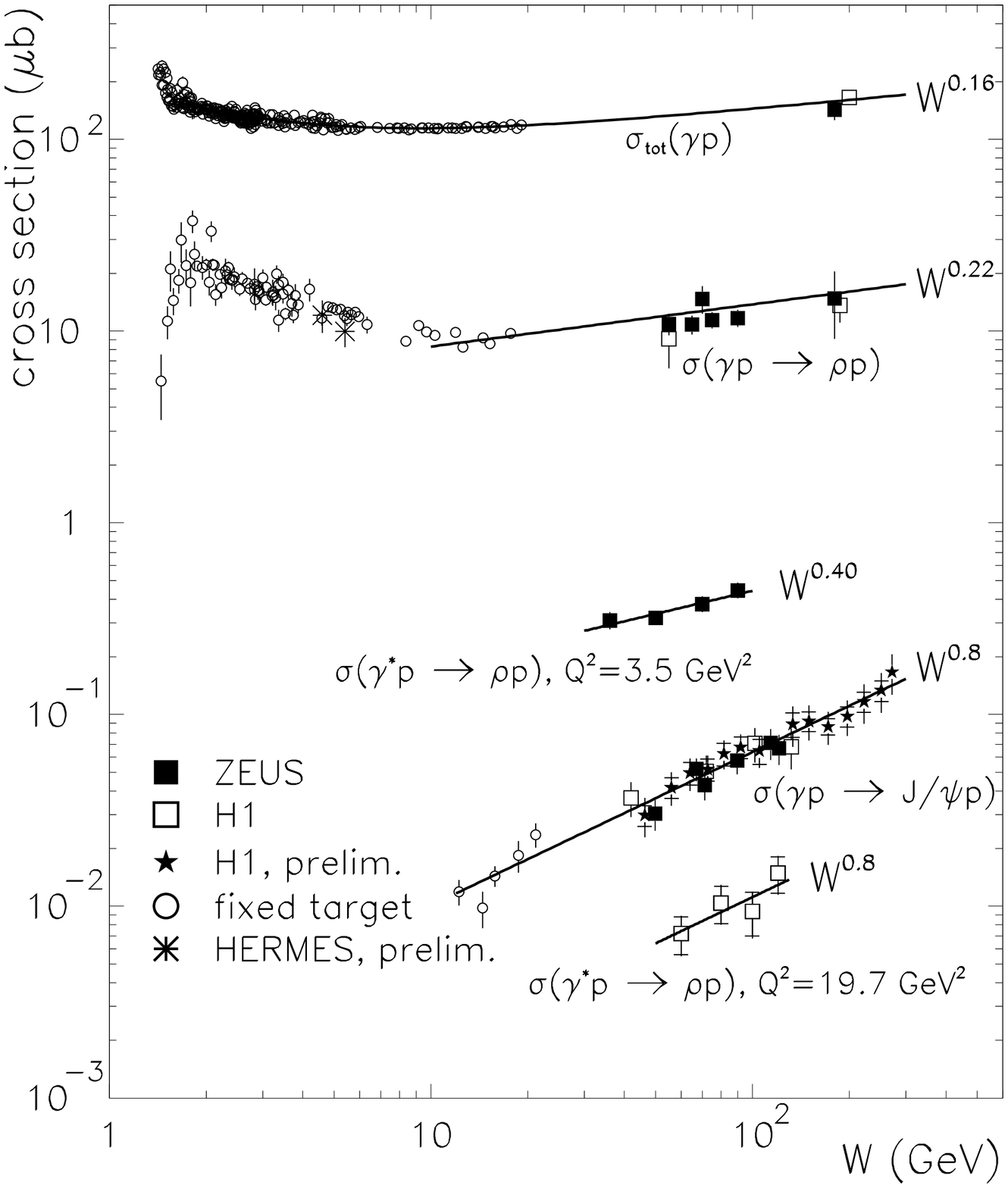,width=0.40\hsize,%
bbllx=17,bblly=98,bburx=534,bbury=708}
\end{center}
\vspace*{-0.5cm}
\caption{Left: $W$ dependence of the cross section 
  $\sigma(\gp \rightarrow Vp)$ for various vector mesons, from fixed
  target and HERA measurements.  Right: $W$ dependence of $\rho^0$
  cross section for different $Q^2$ values compared to $J/\psi$
  photoproduction.}
\label{fig:photo-xsec}
\end{figure}

\begin{figure}[htb]
\begin{center}
\epsfig{file=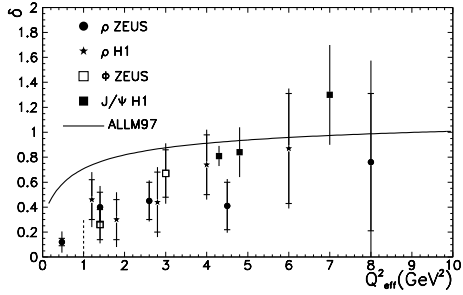,width=0.4\hsize,
bbllx=18,bblly=235,bburx=547,bbury=577}
\end{center}
\vspace*{-0.5cm}
\caption{The power $\delta$ of the $W^{\delta}$ dependence of 
  $\sigma(\gamma^{(\star)}p \rightarrow Vp)$ as a function of the
  effective scale of the scattering, $Q^2_{eff}$. The curve represents
  the expectations from the $W$ dependence of the total $\gvp$ cross
  sections.  }
\label{fig:delta-eff}
\end{figure}

The $W$ dependence of exclusive $V$ photoproduction is shown in
Figure~\ref{fig:photo-xsec}. The striking fact is the fast rise of the
$J/\psi$ cross section, which has now been measured by the H1
experiment up to $W=300 \gev$~\cite{h1-jpsi}. This rise is in good
agreement with perturbative QCD
calculations~\cite{Frankfurt:1996,Frankfurt:1998}.  For $\rho^0$
production~\cite{zeus-rho,h1-rho}, the rise of the cross section with
$W$ depends on $Q^2$, as shown in Figure~\ref{fig:photo-xsec}.  As can
be seen, only at about $Q^2=20 \gevtwo$ the steepness of the rise is
comparable to that of the $J/\psi$.
This may be attributed to the effective scale in $V$ production,
$Q^2_{eff}$, which depends not only on $Q^2$ but also on wave function
effects~\cite{Frankfurt:1996}. The rise with $W$ of $\sigma_V \sim
W^{\delta}$ can be quantified by $\delta$, which is shown in
Figure~\ref{fig:delta-eff} as a function of $Q^2_{eff}$.  The results
for different processes seem to line up along the same curve, which,
above $Q^2_{eff} \simeq 4 \gevtwo$, follows the expectations from the
$W$ dependence of $\gvp$ total cross section.

\begin{figure}[hbt]
\begin{center}
\epsfig{file=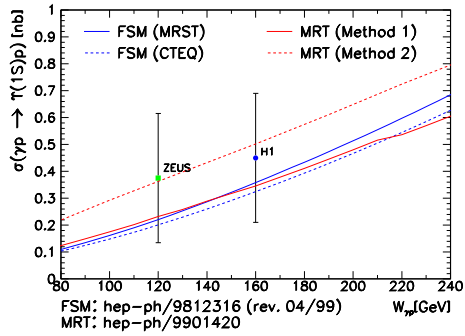,width=0.4\hsize}
\end{center}
\vspace*{-0.5cm}
\caption{The $\Upsilon$ photoproduction cross section as a function of
  $W$ compared to calculations of perturbative QCD,
  FSM~\protect\cite{Frankfurt:1999a}, MRT~\protect\cite{Martin:1999}. }
\label{fig:upsilon-xsec}
\end{figure}
Contrary to expectations, the $\Upsilon$ production cross section was
found to be larger than predicted by perturbative
calculations~\cite{zeus-upsilon,h1-upsilon}. This disagreement led to
two findings~\cite{Frankfurt:1999a,Martin:1999}. The gluons exchanged
in $V$ production do not have the same value of $x$. For a $V$ with
mass $M_V$ produced from a $\gamma^\star$ with virtuality $Q^2$, the
difference $\delta x$ is given by
\begin{equation}
\delta x = \frac{M_V^2+Q^2}{W^2+Q^2} \, .
\label{eq:deltax}
\end{equation}
For $\Upsilon$ production, $\delta x$ can be large and the skewed gluon
distribution has to be used. The skewed (also called off-forward)
parton distributions \index{skewed parton distributions} are hybrid
objects, which combine properties of form factors and ordinary parton
distributions~\cite{Radyushkin:1997,Ji:1998}.  The use of the
skewed distribution leads to an enhancement of $\sigma_{\Upsilon}$ by a 
factor $\simeq 2$. For a process with a fast rise with $W$, the
relation between the elastic cross section and the total cross section
has the form
\begin{equation}
\left.\frac{d\sigma_{el}}{dt}\right|_{t=0}=\left(1+\left(\frac{Re A}{Im A}\right)^2\right)\frac{\sigma_{tot}^2}{16\pi} \, ,
\label{eq:elupsilon}
\end{equation}
where the ratio of the real to imaginary part of the scattering
amplitude, $A$, cannot be neglected.  This produces another
enhancement factor of $1.5 - 1.7$.  The present agreement of
theoretical calculations with data is shown in
Figure~\ref{fig:upsilon-xsec}.

\subsection{$Q^2$ dependence}
\begin{figure}[htb]
\begin{center}
\epsfig{file=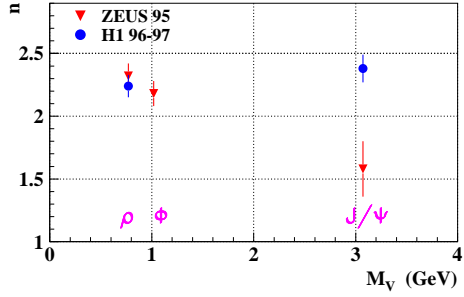,width=0.4\hsize}
\end{center}
\vspace*{-0.5cm}
\caption{The power $n$ in the $(Q^2+M_V^2)^{-n}$ dependence of $
V$ production as a function of mass, $M_V$. }
\label{fig:q2slope}
\end{figure}
The $Q^2$ dependence of the cross section at fixed $W$ (ranging from
75 to 90 GeV) is parameterized as $\sigma_V \sim (Q^2+M_V^2)^{-n}$
\cite{zeus-rho,h1-rho,h1-charmonium,zeus-phi}. The power $n$ as a
function of $M_V$ is shown in Figure~\ref{fig:q2slope} and is found to
be substantially smaller than 3 (see equation~(\ref{eq:sigmal})).

\subsection{$t$ dependence}
\begin{figure}[htb]
\begin{center}
\epsfig{file=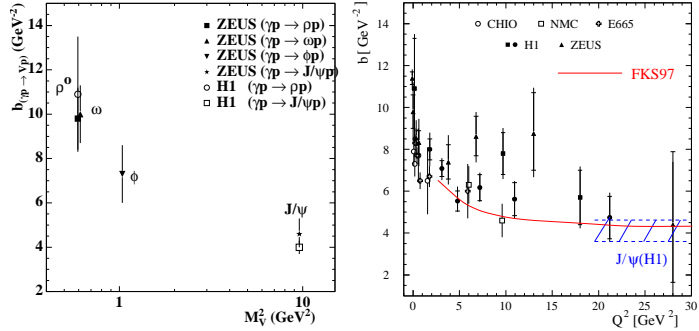,width=0.30\hsize}
\epsfig{file=figs/bslope-fsk.eps,width=0.30\hsize}
\end{center}
\vspace*{-0.7cm}
\caption{Left: the exponential slope $b$ of the $t$ distribution 
  in $\gvp \rightarrow Vp$ at HERA as a function of $M_V^2$.  Right:
  the exponential slope $b$ as a function of $Q^2$ for $\rho^0$
  production.  }
\label{fig:bslope}
\vspace*{-0.1cm}
\end{figure}
\begin{figure}[hbt]
\begin{center}
\epsfig{file=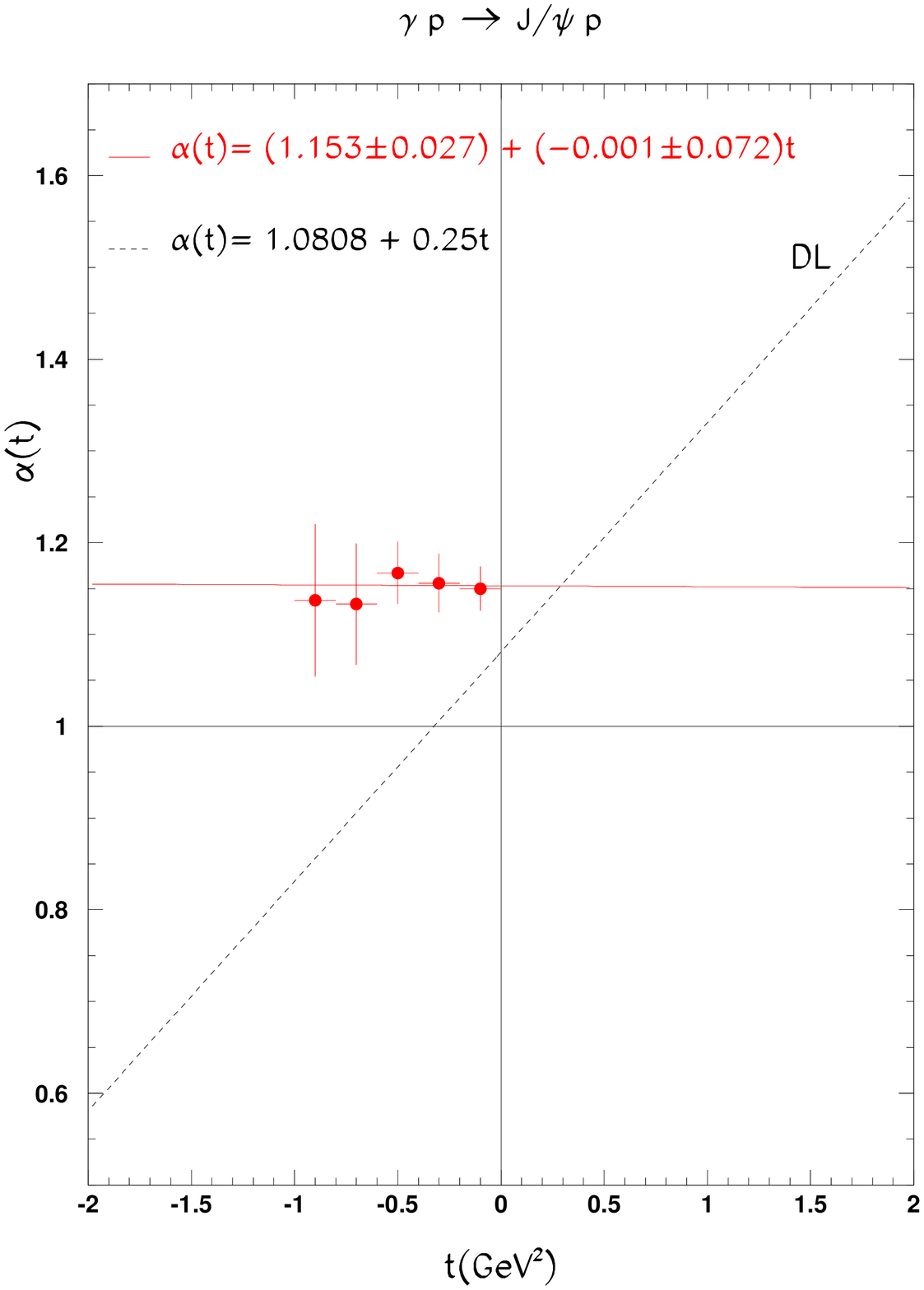,width=0.33\hsize}
\epsfig{file=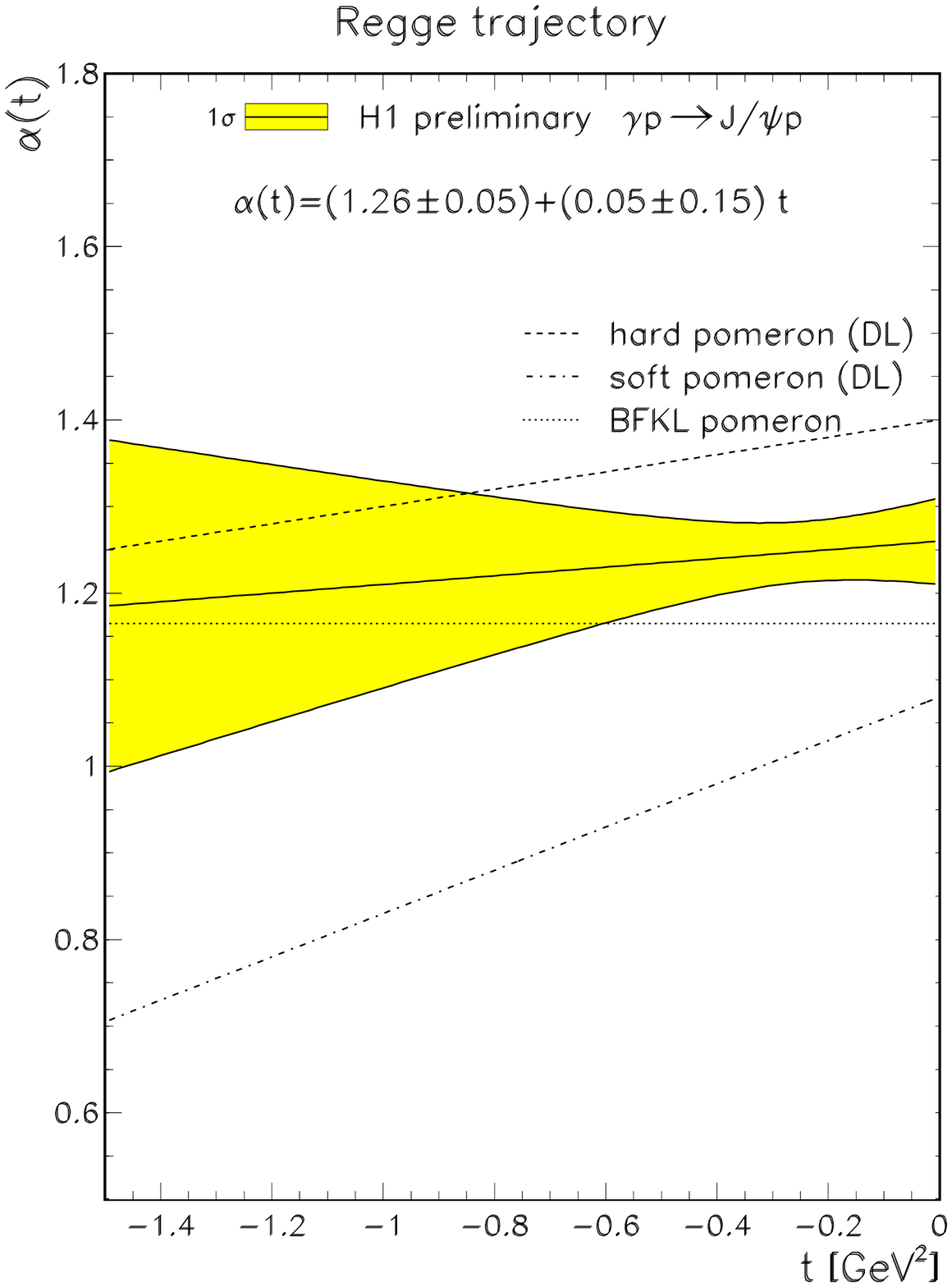,width=0.33\hsize}
\end{center}
\vspace*{-0.7cm}
\caption{The Pomeron trajectory $\apom(t)$ as determined from low energy 
and HERA data (left) and from H1 data alone (right).  }
\label{fig:pomtraj}
\end{figure}

A compilation of the exponential slopes $b$ of the $t$ distributions
at $Q^2\simeq 0$ as a function of $M_V$ is shown in
Figure~\ref{fig:bslope}. For $\rho^0$ production, $b$ is also shown as
a function of $Q^2$. The value of the slope decreases both with the
mass of the $V$ and with $Q^2$, towards the value measured for the
$J/\psi$, pointing to a decreasing size of the interaction region.
The perturbative character of $J/\psi$ production can be best
established by deriving the value of $\aprime$ from the $W$
dependence of the cross section at fixed $t$ (see
equation~(\ref{eq:sel}))
\begin{equation}
\frac{d\sigma}{dt} \sim W^{4(\apom(t)-1)} \, .
\end{equation}
Earlier measurements of HERA, combined with low energy data, indicated
that $\alpha(t) \simeq \mathrm{const}$~\cite{Levy:1998} (see
Figure~\ref{fig:pomtraj} left).  H1 determined $\aprime$ using only
their own data~\cite{h1-jpsi}. The result, shown in
Figure~\ref{fig:pomtraj} with $\aprime = 0.05 \pm 0.15 \gevmtwo$,
supports the hard nature of exclusive $J/\psi$ production at HERA.

\section{Deeply Virtual Compton scattering}

\begin{figure}[htb]
\begin{center}
\epsfig{file=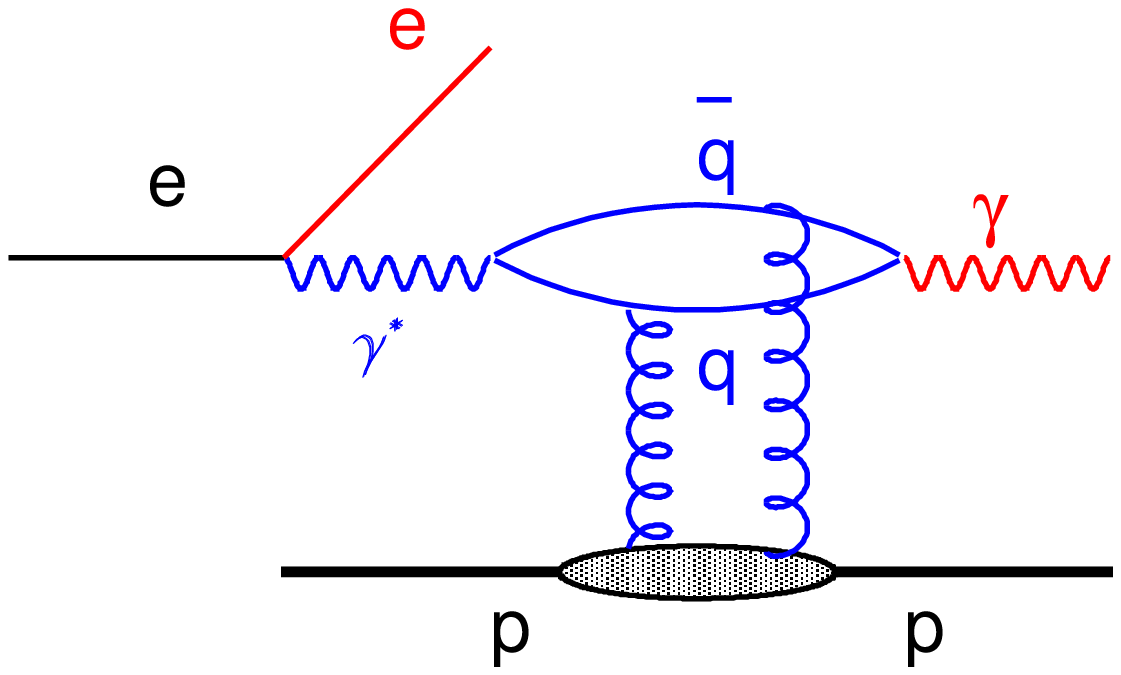,width=0.3\hsize}
\epsfig{file=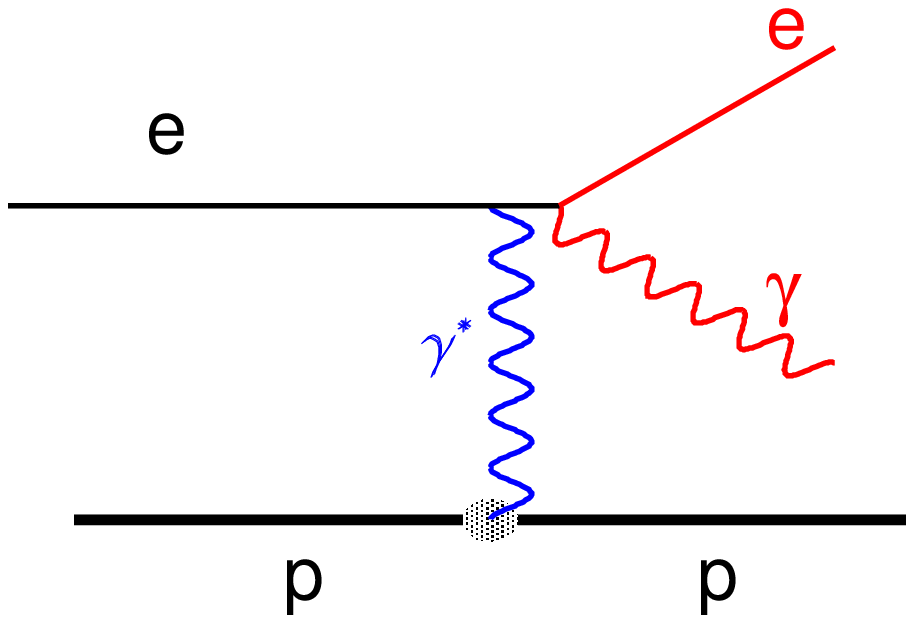,width=0.3\hsize}
\epsfig{file=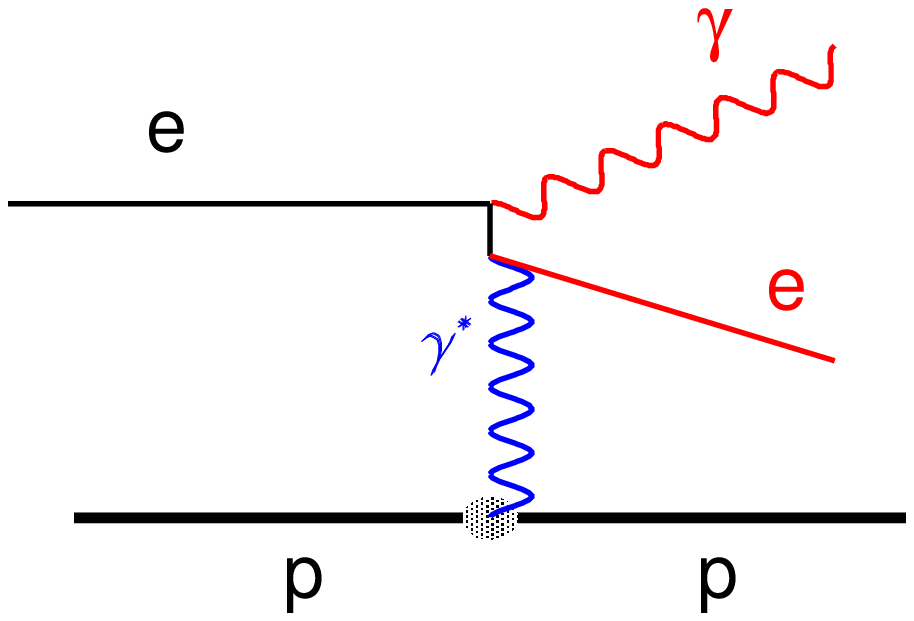,width=0.3\hsize}
\end{center}
\caption{Diagrams for real $\gamma$ production in DIS.  }
\label{fig:dvcs-diag}
\end{figure}
Another example of hard diffractive scattering is the exclusive
production of a real photon in DIS~\cite{Frankfurt:1998a} (see
Figure~\ref{fig:dvcs-diag}).  In this deeply virtual Compton
scattering (DVCS) \index{deeply virtual Compton scattering} (for
recent discussions see~\cite{Radyushkin:1997,Ji:1998}), the cross
section is less suppressed by factor $Q^2$ than in exclusive vector
meson production~\cite{Frankfurt:1998a}, and the interpretation of the
data is easier due to the lack of $V$ wave function effects. DVCS
interferes with the QED Compton process, which leads to an identical
final state, as depicted in Figure~\ref{fig:dvcs-diag}.  The
interference may allow a direct measurement of the real part of the QCD
scattering amplitude.  By extracting the DVCS, 
skewed parton distributions can be measured and additional
information on the $F_2$ of the proton at low
$x$ can be obtained~\cite{Frankfurt:1998a}.
\begin{figure}[htb]
\begin{center}
\epsfig{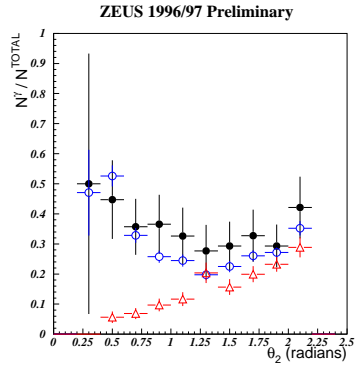}
\end{center}
\vspace*{-0.5cm}
\caption{Fraction of photon candidates, $N^\gamma/N^{\mathrm TOTAL}$, 
  for the second electromagnetic candidate in the calorimeter with
  energy $E>5 \gev$, produced at angle $\theta_2$, closest to the
  proton direction.  The data (dots) are compared to MC expectations
  for GenDVCS (circles) and for elastic Compton scattering only
  (triangles).  }
\label{fig:dvcs-signal}
\end{figure}

The ZEUS experiment has searched for DVCS in DIS events with $Q^2>6
\gevtwo$ and for $5\cdot 10^{-4}<x<10^{-2}$~\cite{zeus-dvcs}. Events
with two electromagnetic, $e/\gamma$, candidates were selected with no
additional hadronic activity in the central detector. The second
candidate was defined as the one produced closest to the original
proton direction, with angle $\theta_2$, and was required to be in the
region of the detector where tracking is available. The data were
compared to the expectations of a QED Compton MC~\cite{qedc-mc} and a
MC which included DVCS (GenDVCS)~\cite{dvcs-mc}. The results are shown
in Figure~\ref{fig:dvcs-signal}. A clear excess of $\gamma$ candidates
at low $\theta_2$ is observed over expectations from the QED Compton
process, while their rate is well reproduced by GenDVCS. The signal
for DVCS is now established at HERA.

\section{Outlook}

This report is far from a complete account of diffractive physics.
Its main purpose was to review the progress achieved in understanding
diffractive phenomena with the language of perturbative QCD.  Much has
been learned from studying inclusive diffraction, where the gluons are
found to be the main players. Diffractive DIS provides a complementary
area to inclusive DIS in studying the dynamics of QCD at high
energy. The new class of hard diffractive processes, such as exclusive
vector meson production and deeply virtual Compton scattering, fulfill
the properties expected from perturbative QCD. A comprehensive study
of these processes may provide, in the future, important information
on the gluon content in the proton and on the wave functions of vector
mesons. The accumulated data cannot be described coherently by one
universal Pomeron trajectory. The issue of what the Pomeron is
requires further study.

Two subjects of potential interest, which were not discussed due to
time constraints, are the large $t$ diffractive scattering studied at
FNAL~\cite{cdf-lrgjets,d0-lrgjets} and HERA~\cite{h1-hight} and the
double-$\pom$ exchange studied at FNAL~\cite{cdf-dpe}. Here more
theoretical work is necessary.  It is quite clear that the emerging
picture of diffraction will not be satisfactory without a complete
account of all hard diffractive processes.


\bigskip I would like to thank all my colleagues who have helped me in
preparing this report. My special gratitude goes to M.  Albrow, J.
Bartels, K. Borras, A. Brandt, J. Dainton, K. Goulianos, L.
Frankfurt, H. Kowalski, A. Levy, M. McDermott, N. Makins, K.  Mauritz
and M. Strikman who patiently discussed the data
and their interpretation with me. 


\def\Discussion{
\setlength{\parskip}{0.3cm}\setlength{\parindent}{0.0cm}
     \bigskip\bigskip      {\Large {\bf Discussion}} \bigskip}
\def\speaker#1{{\bf #1:}\ }
 
\Discussion
 
\speaker{Harry Lipkin (Weizmann Institute)} I would like to point out
that the Donnachie and Landshoff  fit to the total cross section,
which was accepted by the Particle Data Group in 1966, was thrown out
by them in 1998 with a different value of the pomeron intercept.  I
have fit the same data some time ago with a still different value.
 The point is that there are still not enough data on total cross
sections to tell us whether there is a universal pomeron with a
 universal intercept that is coupled to meson-baryon and baryon-baryon
total cross sections.    What are needed are new data at higher
energies.  The recent data on hyperon-nucleus cross sections from
SELEX actually support my model, but I don't believe that either.    I
think that what is happening is that there is a non-leading
contribution which confuses the issue and is not  understood. To
resolve this, we just have to get more data at higher energies.
 
\noindent
Answer: There is no denying that a well ``measured'' soft Pomeron
would help to establish when the departure from universality becomes
effective. However, it is difficult to imagine that it would alter in
any way the fact that the $W$ dependence of the total $\gvp$ cross
section changes with $Q^2$ and follows the laws of QCD evolution. I
think that even the diffractive data, imprecise as they are, cannot
be accommodated by a universal trajectory. It may well be that the
idealized Pomeron of Gribov is just a mathematical construction, while
the pomeron we try to investigate is just another name for the nature
of strong interactions. The latter, by virtue of QCD and the interplay of
soft and hard physics, remain versatile.

\end{document}